\documentclass[twocolumn,twocolappendix]{aastex631}
\usepackage{amsmath,amssymb}
\usepackage{booktabs}
\usepackage{xspace}
\usepackage{needspace}

\shorttitle{The Hexadecapole Nearly Cancels the Octupole}
\shortauthors{Klein}

\defcitealias{katz2011}{KDM2011}
\defcitealias{klein2024a}{Paper I}
\defcitealias{klein2024b}{Paper II}
\defcitealias{klein2024c}{Paper III}
\defcitealias{klein2026}{Paper IV}
\newcommand{\citekdm}{\citetalias{katz2011}\xspace}
\newcommand{\citepaperI}{\citetalias{klein2024a}\xspace}
\newcommand{\citepaperII}{\citetalias{klein2024b}\xspace}
\newcommand{\citepaperIII}{\citetalias{klein2024c}\xspace}
\newcommand{\citepaperIV}{\citetalias{klein2026}\xspace}
\newcommand{\Jh}{\hat{J}}
\newcommand{\chiH}{\chi_{\rm H}}
\newcommand{\epsh}{\epsilon_{\text{hex}}}
\newcommand{\epstwo}{\epsilon_{2}}
\newcommand{\eper}{e_{\text{per}}}

\makeatletter
\AtBeginDocument{}
\makeatother

\begin{document}

\title{Four Which I Knew Not:\\
The Hexadecapole Nearly Cancels the Octupole\\
in Librating Kozai--Lidov Cycles}

\correspondingauthor{Ygal Y. Klein}
\email{ygalklein@gmail.com}

\author[0009-0004-1914-5821]{Ygal Y. Klein}
\affiliation{Institute for Advanced Study,
Einstein Drive, Princeton, NJ 08540, USA}

\begin{abstract}
Phenomena requiring close proximity between two bodies, from
black-hole mergers to hot Jupiters, may originate in
hierarchical triples, in which secular torques from a distant companion
drive the inner binary's eccentricity nearly to unity. Population syntheses of these channels commonly
integrate the double-averaged (DA) secular equations truncated at
octupole order. We show that, for a fixed outer orbit, this truncation
fails for the slow evolution on librating Kozai--Lidov cycles, in which the argument of
pericenter librates. On these cycles the leading octupole kicks to
$j_z$, the component of the inner angular momentum along the outer one,
cancel pairwise, and the hexadecapole enters at the order of the
surviving octupole torque, with the opposite sign. Strikingly, as the
inner mass ratio vanishes the two nearly cancel, and their sum changes
sign within the librating window, so octupole-truncated integrations
omit part of the leading slow torque on this class. For
$\left|j_z\right|\ll1$, the slow dynamics driven by the
hexadecapole is approximately solved analytically as a simple pendulum
in the doubled azimuth of the eccentricity vector, with closed-form
coefficients on rotating and librating cycles. For a circular outer
orbit the hexadecapole-modified cycles are solved exactly. For unequal
inner masses the octupole adds to the pendulum's torque on librating cycles,
where its alternating kick can also flip orbits, and generally
dominates rotating ones. DA integrations follow the predicted
excursions, periods and flip criteria. Population consequences depend
on the validity of the high-eccentricity reduction and on other
processes affecting the cycles.
\end{abstract}

\keywords{Celestial mechanics (211) --- Three-body problem (1695) ---
Multiple stars (1081) --- Exoplanet dynamics (490)}

\section{Introduction}
\label{sec:intro}

Many astrophysical phenomena require two bodies to come into close
proximity. In a hierarchical triple, in which an inner binary is
perturbed by a distant companion, secular torques from the companion can drive
large oscillations of the inner eccentricity and inclination on
timescales long compared with both orbital periods, and near the
eccentricity maximum the two inner bodies pass close to each other.
The gravitational perturbation is expanded in the ratio of the inner
to outer semimajor axes. At leading (quadrupole) order, the secular
equations for a fixed outer orbit admit periodic eccentricity and
inclination oscillations, the Kozai--Lidov Cycles (KLCs;
\citealt{kozai1962,lidov1962}).\footnote{For earlier work by
\citet{vonzeipel1910}, see the historical discussion in
\citet{ito2019}.} For an eccentric perturber, the next (octupole)
order produces extreme eccentricities and orbital flips, changes between
prograde and retrograde motion relative to the outer orbit (Eccentric
Kozai--Lidov, EKL;
\citealt{ford2000,naoz2011,lithwick2011,sidorenko2018,lei2022}; for a
review see
\citealt{naoz2016}). The following (hexadecapole) order is the
subject of this Letter.

Secular eccentricity excitation in hierarchical triples has been
invoked for hot-Jupiter formation and the eccentricity distribution of cold Jupiters
\citep{naoz2011,naoz2012,petrovich2015,anderson2016,munoz2016,
vick2019,oconnor2021,stephan2021,
angelo2022,stephan2024wdkick,weldon2025,grishin2026},
binary black-hole mergers
\citep{miller2002,blaes2002,wen2003,antonini2012,naoz2013pn,
antonini2016,liu2017,hoang2018,grishin2018,liu2018,
hamilton2019a,liu2019quad,martinez2020,fragione2019,vignagomez2025},
and neutron-star gravitational-wave sources \citep{feng2026}.
Other applications include blue stragglers
\citep{perets2009,naoz2014,antonini2016}, tidal disruption events
\citep{ivanov2005,melchor2024,melchor2025}, Type Ia supernovae
\citep{thompson2011,katz2012,haim2018}, white-dwarf binaries, cataclysmic
variables, and low-mass X-ray binaries
\citep{shariat2023wd,shariat2025cv,shariat2025lmxb}, fast-radio-burst
progenitors \citep{shariat2026frb}, and Kuiper Belt contact binaries
such as Arrokoth \citep{grishin2020}.
Averaging the equations of motion over both orbital periods gives the
double-averaged (DA) equations, and population calculations often
integrate them truncated at the octupole order
\citep[e.g.][]{naoz2012,stephan2016,hoang2018,
shariat2023wd,weldon2025,shariat2026frb}.

KLCs come in two classes, distinguished by the behavior of the
argument of pericenter of the inner orbit: it circulates for
\textit{rotating} cycles and oscillates about a fixed value for
\textit{librating} cycles. The quadrupole KLC is itself exactly a
simple pendulum \citep{basha2025}. Higher multipoles modulate these
cycles slowly, and this slow evolution has been described,
approximately, by simple pendulums as well. For rotating cycles at
high eccentricities, \citet[hereafter \citekdm]{katz2011} obtained
equations averaged over the KLC for the slow evolution of $j_z$, the
component of the inner binary's dimensionless angular momentum along
the outer angular momentum, and $\Omega_e$, the azimuth of the
eccentricity vector, which points toward the inner pericenter; both
are defined explicitly in Section \ref{sec:setup}.
\citet[hereafter \citepaperI]{klein2024a} showed that for most
initial conditions these slow equations reduce to those of a simple
pendulum. \citet[hereafter \citepaperII]{klein2024b} showed that the
slow evolution remains a simple pendulum when the correction to double
averaging described by Brown's Hamiltonian
\citep{brown1936a,brown1936b,brown1936c,soderhjelm1975,
cuk2004,breiter2015,luo2016,will2021,tremaine2023,grishin2024a,
lei2025a,lei2025b,gao2025} is included, and
\citet[hereafter \citepaperIII]{klein2024c} obtained a pendulum
description of the resonance with a precessing quadrupole
\citep{hamers2017,petrovich2017,klein2023,klein2024d}.

On rotating cycles the azimuth of the eccentricity vector precesses
slowly and continuously, so the leading octupole changes in $j_z$
accumulated over each cycle (the per-cycle kicks) add up from one
cycle to the next and drive the pendulum of \citepaperI. On librating
cycles at $j_z=0$, instead, the eccentricity vector crosses the
axis of the outer angular momentum once during each cycle, so
its azimuth advances by $\pi$, and the kicks alternate in sign and
cancel pairwise (\citekdm).
\citekdm\ noted that librating cycles can still be modulated, and can
flip; \citet[hereafter \citepaperIV]{klein2026} showed that this remaining
slow evolution is again a simple pendulum, now in the doubled azimuth
of the eccentricity vector, with a torque quadratic in the
octupole strength. Because the contribution linear in that
strength cancels, the next term in the gravitational expansion,
the hexadecapole, may be needed for a consistent leading-order
description. For equal inner masses, where the octupole vanishes,
\citet{will2017} found numerically that the hexadecapole can produce
flips, in integrations in which the outer orbit also evolves. The DA equations have been derived through hexadecapole order
\citep{will2017,hamers2016}, and recently the hexadecapole has been
included in applications to black-hole mergers and their
electromagnetic counterparts \citep{naoz2025triples} and to the
misalignment of the VHS~1256$-$1257 system \citep{holzknecht2026}.

In this Letter we obtain an approximate analytic solution for the
slow high-eccentricity evolution of $j_z$ and $\Omega_e$ through
hexadecapole order, assuming a fixed outer orbit and
$\left|j_z\right|\ll1$. The leading
slow equations reduce to a simple pendulum in the doubled azimuth
of the eccentricity vector on both KLC classes, with coefficients
obtained in closed form. For equal inner masses this supplies the
leading slow evolution of $j_z$ (Section \ref{sec:equalmass}). We
then combine it with the octupole for unequal inner masses $m_1$
and $m_2$ (Section
\ref{sec:competition}). On the librating class the two torques
enter at the same order in the ratio of the semimajor axes and
oppose one another. In the limit of a small inner mass ratio,
$m_2/m_1\to0$, the two contributions nearly cancel, and their sum
changes sign within the librating window. Integrations of the DA
equations truncated at the octupole order therefore omit part of the
leading slow torque on the librating class. For unequal masses we
also derive a flip criterion that includes the flips caused by the
alternating octupole kick.
We compare the analytic predictions with numerical integrations of
the DA equations and examine the consequences for a planetary
system. For a circular outer orbit, where only the axisymmetric
hexadecapole remains, the modified cycles are solved exactly by a
single integral (Appendix \ref{sec:circular}).

\section{Double-averaged equations}
\label{sec:setup}

We consider a hierarchical triple: an inner binary of masses $m_1$
and $m_2$ (mass ratio $q\equiv m_2/m_1\leq1$), semimajor axis $a$ and
eccentricity $e$, and a distant perturber of mass
$m_\text{per}$ on an orbit of semimajor axis $a_\text{per}$ and
eccentricity $\eper$, with $\alpha\equiv a/a_\text{per}\ll1$. We neglect the
back-reaction on the outer orbit: its angular
momentum, which we take to be aligned with the conserved total angular
momentum, and its eccentricity vector are held fixed, and they define
the axes: $\hat{\mathbf{z}}$ along the outer angular momentum and
$\hat{\mathbf{x}}$ toward the outer pericenter. This approximation
requires their evolution to be negligible over the times considered.
The inner binary precesses the outer pericenter at a rate proportional
to the ratio of the inner to the outer circular orbital angular
momentum, $L_\text{in}/L_\text{out}$, and the approximation holds while this
precession is slow compared with the slow evolution of
Sections \ref{sec:equalmass} and \ref{sec:competition}. This does not
require $q\to0$: a perturber much more massive than the inner binary
also makes $L_\text{in}/L_\text{out}$ small, and we use the
approximation for equal inner masses as well
(Section \ref{sec:equalmass}).

We work in the secular approximation. Averaging over the inner
orbital period keeps $a$ constant, and the inner orbit is then
described by its eccentricity vector $\mathbf{e}$ and its
dimensionless angular momentum
$\mathbf{j}=\mathbf{J}_1/\big(\mu_{12}\sqrt{GM_{12}a}\big)$, where
$\mathbf{J}_1$ is the orbital angular momentum of the inner binary,
$M_{12}=m_1+m_2$ and $\mu_{12}=m_1m_2/M_{12}$. The two vectors satisfy
$\mathbf{j}\cdot\mathbf{e}=0$ and $j^2+e^2=1$. The disturbing function
per unit inner reduced mass is expanded in $\alpha$ through
hexadecapole order and averaged over both orbits:
\begin{equation}
\begin{gathered}
\Phi_\text{per}=\Phi_{0}\left(\phi_\text{quad}
+\epsilon_\text{oct}\,\phi_\text{oct}
+\epsh\,\phi_\text{hex}\right),\\
\Phi_{0}=\frac{Gm_\text{per}a^{2}}{a_\text{per}^{3}
\left(1-\eper^{2}\right)^{\frac{3}{2}}},
\end{gathered}
\label{eq:phi_Per}
\end{equation}
with \citep[e.g.][]{katz2011,naoz2016,will2017}
\begin{equation}
\phi_\text{quad}=\frac{3}{4}\left(\frac{1}{2} j_{z}^{2}+e^{2}
-\frac{5}{2} e_{z}^{2}-\frac{1}{6}\right),
\label{eq:phi_quad}
\end{equation}
\begin{equation}
\phi_\text{oct}=\frac{75}{64}\left(e_{x}\left(\frac{1}{5}
-\frac{8}{5} e^{2}+7e_{z}^{2}-j_{z}^{2}\right)
-2e_{z}j_{x}j_{z}\right).
\label{eq:phi_oct}
\end{equation}
The hexadecapole term is the new ingredient here, compared with
\citekdm, Papers I--IV and the octupole-level EKL studies
\citep[e.g.][]{ford2000,lithwick2011,naoz2016}. Its
contribution to
the equations of motion was given in orbital-element form by
\citet{will2017} (see also \citealt{hamers2016}); in vector
form the term splits into an axisymmetric and a nonaxisymmetric part,
\begin{equation}
\phi_\text{hex}=\left(1+\frac{3}{2}\eper^2\right)
f_0\left(\mathbf{e},\mathbf{j}\right)
+\eper^2\,f_2\left(\mathbf{e},\mathbf{j}\right),
\label{eq:phihex}
\end{equation}
where $f_0$ and $f_2$ are polynomials in the components of
$\mathbf{e}$ and $\mathbf{j}$, listed in Table \ref{tab:hex}
(Appendix \ref{sec:coefficients}). The strengths are
\begin{equation}
\begin{gathered}
\epsilon_\text{oct}=\alpha\,\frac{\eper}{1-\eper^{2}}\,
\frac{1-q}{1+q},\\
\epsh=\alpha^{2}\,\frac{1}{\left(1-\eper^{2}\right)^{2}}\,
\frac{1+q^{3}}{\left(1+q\right)^{3}} ,
\end{gathered}
\label{eq:strengths}
\end{equation}
where the $q$-dependent factors are the mass factors
$\big(m_1^{\ell-1}+(-1)^{\ell}m_2^{\ell-1}\big)/M_{12}^{\ell-1}$ of the
multipole of order $\ell$ ($\ell=2,3,4$ for the quadrupole, octupole and
hexadecapole). Every odd multipole therefore vanishes identically at
$q=1$, while the hexadecapole factor only drops to $\frac{1}{4}$.

At $\eper=0$ the averaged outer orbit is a circular ring, so the
interaction is axisymmetric at every multipole order: the octupole
and the $f_2$ term, which carry factors of $\eper$ and $\eper^2$
(Equations \ref{eq:phihex} and \ref{eq:strengths}), vanish, and only
the $f_0$ term survives. The axisymmetric term $f_0$ conserves $j_z$
and only modifies the KLCs themselves, their shape and period.
With $j_z$ conserved, the motion at $\eper=0$ has one degree of
freedom and is solved exactly by a single integral
(Appendix \ref{sec:circular}).

For $\eper>0$ the $f_2$ term enters as well. Like the octupole, it
depends on the direction of the outer pericenter, but through a
different azimuthal harmonic: a rotation of the inner orbit about
$\hat{\mathbf{z}}$ by an angle $\varphi$ changes $\phi_\text{oct}$
through $\cos\varphi$ and $\sin\varphi$ (the first harmonic, $m=1$)
and $f_2$ only through $\cos2\varphi$ and $\sin2\varphi$ (the second,
$m=2$). The perturber moves in the \textit{x-y} plane, and the
hexadecapole contains the direction to it only in even powers, so only
the harmonics $m=0$, $2$ and $4$ can occur. The $m=4$ harmonic
vanishes, because the outer averaging weight
$\left(1+\eper\cos f\right)^3$, where $f$ is the outer true anomaly,
is cubic in $\cos f$. Its mean, $1+\frac{3}{2}\eper^2$, and its
$\cos2f$ component, proportional to $\eper^2$, give the two factors of
Equation \ref{eq:phihex}, which therefore contains the full dependence
of the normalized hexadecapole on the outer eccentricity.

We denote the strength of the nonaxisymmetric term by
$\epstwo\equiv\epsh\eper^2$. From Equation \ref{eq:strengths},
\begin{equation}
\epstwo\left(q{=}0\right)=\epsilon^2_\text{oct}\left(q{=}0\right).
\label{eq:bookkeeping}
\end{equation}
Thus the hexadecapole torque is of the same hierarchy order,
$O\left(\alpha^2\right)$, as a second-order octupole torque. For
general $q<1$ the relation reads
$\epstwo=\epsilon^2_\text{oct}\left(1+q^3\right)/\big[\left(1+q\right)\left(1-q\right)^2\big]$,
with a factor that depends on $q$ alone. At $q=1$, where
$\epsilon_\text{oct}=0$, Equation \ref{eq:strengths} gives directly
\begin{equation}
\epstwo\left(q{=}1\right)=\frac{\alpha^2\eper^2}{4\left(1-\eper^2\right)^2}.
\label{eq:eps2q1}
\end{equation}

The equations
of motion are the standard Milankovitch pair
\citep[e.g.][]{milankovitch1939,tremaine2009}, which we integrate
numerically:
\begin{align}
\frac{\mathrm{d}\mathbf{j}}{\mathrm{d}\tau}&=
\mathbf{j}\times\nabla_{\mathbf{j}}\phi
+\mathbf{e}\times\nabla_{\mathbf{e}}\phi,
\label{eq:milankovitch_j}\\
\frac{\mathrm{d}\mathbf{e}}{\mathrm{d}\tau}&=
\mathbf{j}\times\nabla_{\mathbf{e}}\phi
+\mathbf{e}\times\nabla_{\mathbf{j}}\phi ,
\label{eq:milankovitch_e}
\end{align}
with $\phi\equiv\Phi_\text{per}/\Phi_0$ and time
measured in units of the secular timescale
$P_\text{sec}=\sqrt{GM_{12}a}/\Phi_0$, $\tau=t/P_\text{sec}$.

At quadrupole order ($\epsilon_\text{oct}=\epsh=0$) the interaction is
axisymmetric about $\hat{\mathbf{z}}$ for any $\eper$, so $j_z$ is
conserved; since $\phi_\text{quad}$ is conserved as well, so is the
Kozai constant
\begin{equation}
\begin{aligned}
C_{K}&=\frac{4}{3}\phi_\text{quad}+\frac{1}{6}-\frac{1}{2}j_{z}^{2}\\
&=e^{2}-\frac{5}{2}\,e_{z}^{2}
=e^{2}\left(1-\frac{5}{2}\sin^{2}i\,\sin^{2}\omega\right),
\end{aligned}
\label{eq:CK}
\end{equation}
with $-\frac{3}{2}\leq C_K\leq1$, where $i$ is the inclination and
$\omega$ the argument of pericenter. The last form separates the two
KLC classes: wherever $\sin\omega=0$ it gives $C_K=e^2\geq0$, so
cycles in which $\omega$ circulates, crossing $\sin\omega=0$ every
cycle, have $C_K>0$ (the \textit{rotating} class), while cycles with
$C_K<0$ never reach $\sin\omega=0$ and their $\omega$ librates about
$\pm\frac{\pi}{2}$ (the \textit{librating} class). As in
\citekdm\ and Papers I--IV we focus on the
high-eccentricity regime $\left|j_z\right|\ll1$. Denoting the polar angle and the azimuth of the eccentricity vector
by $i_e$ and $\Omega_e$, as in \citekdm, we have
\begin{equation}
\mathbf{e}=e\left(\sin i_e\cos\Omega_e,\,\sin i_e\sin\Omega_e,\,
\cos i_e\right),
\label{eq:evec}
\end{equation}
with the azimuth $\Omega_e$ measured from $\hat{\mathbf{x}}$
(Appendix \ref{sec:angles} relates $i_e$ and $\Omega_e$ to the orbital
elements; Figure \ref{fig:angles}). At quadrupole order the azimuth evolves, exactly at every $j_z$, as
(\citekdm)
\begin{equation}
\frac{\mathrm{d}\Omega_e}{\mathrm{d}\tau}=f_\Omega\,j_z,\qquad
f_\Omega=\frac{3}{4}\left(4-\frac{3}{\sin^{2}i_e}\right).
\label{eq:fOmega}
\end{equation}
With $\left|j_z\right|\ll1$ the azimuth is therefore a slow variable.
For $j_z=0$ it is constant on rotating cycles, on which $\mathbf{e}$
never points along $\hat{\mathbf{z}}$. On librating cycles it is
constant only modulo $\pi$: once per cycle $\mathbf{e}$ crosses
$\hat{\mathbf{z}}$, where the azimuth is undefined and
Equation \ref{eq:fOmega} is indeterminate, and $\Omega_e$ advances by
$\pi$, the half-turn of Equation \ref{eq:halfturn}.

\section{Equal-mass inner binaries}
\label{sec:equalmass}

At $q=1$ every odd multipole vanishes
(Equation \ref{eq:strengths}), so $\epsilon_\text{oct}=0$ for all
$\eper$. For $\eper>0$ both hexadecapole parts survive (mass factor
$\frac{1}{4}$), and
the leading torque on $j_z$ is supplied by the $f_2$ term, of strength
$\epstwo\left(q{=}1\right)$ (Equation \ref{eq:eps2q1}). As for the octupole torque (Papers I
and IV), the way this torque accumulates differs between the two KLC
classes: the azimuth precesses slowly and continuously on the rotating
side and advances by an exact half-turn per cycle on the librating
side. After the common ground of Sections
\ref{sec:cycles}--\ref{sec:kick}, we treat each class in its own
subsection (Sections \ref{sec:rotclass} and \ref{sec:pendulum}).

\subsection{Quadrupole cycles at \texorpdfstring{$j_z=0$}{jz=0}}
\label{sec:cycles}

This subsection is a reminder of the quadrupole cycles at $j_z=0$. It
follows appendix A of \citepaperIII, a
special case of the simple pendulum that \citet{basha2025} obtained
at every $j_z$, with the per-cycle quantities of \citekdm\ and
\citepaperIV. We take the limit $\epsh=0$, in which the cycles are
exact KLCs, at $j_z=0$. The vector $\mathbf{j}$ then oscillates along
a fixed line in the \textit{x-y} plane and we may align the
\textit{y} axis with it, so that $j_x=j_z=e_y=0$ at all times and
$C_K=e_x^2-\frac{3}{2}e_z^2$ (the frame of \citepaperIV\ and, at
$j_z=0$, of \citepaperIII). The in-plane dynamics
$\left(e_x,j_y,e_z\right)$ are then an exact simple pendulum, whose
libration is a librating KLC and whose rotation is a rotating KLC.
With
\begin{equation}
C_x=\frac{9}{8}\left(2C_K+3\right),\qquad
C_z=\frac{27}{8}\left(1-C_K\right),
\label{eq:CxCzk}
\end{equation}
the parameter $k=C_x/C_z$ satisfies $k<1$ on the librating class and
$k>1$ on the rotating class, and the eccentricity cycle, from one
$e\rightarrow1$ passage to the next, is solved by Jacobi elliptic
functions of parameter $k$ on librating cycles and of parameter $1/k$
on rotating ones. Its period is, respectively,
\begin{equation}
T_e=\frac{2K\left(k\right)}{\sqrt{C_z}}
\quad\text{and}\quad
T_e=\frac{2K\left(1/k\right)}{\sqrt{C_x}},
\label{eq:Te}
\end{equation}
where $K$ is the complete elliptic integral of the first kind as a
function of its parameter (the full $\mathbf{j}$-line oscillation
takes $2T_e$, cf.\ \citepaperIII).

The property that separates the classes is the \textit{half-turn} of
\citekdm\ and \citepaperIV:
$\mathbf{e}$ is confined to the plane
perpendicular to the $\mathbf{j}$-line, so $\Omega_e$ is constant
during a cycle unless $\mathbf{e}$ crosses $\hat{\mathbf{z}}$, where
Equation \ref{eq:fOmega} is indeterminate, which
requires $e=\left|e_z\right|$ and is therefore possible only for $C_K<0$. For
librating cycles this crossing happens once per eccentricity cycle:
the in-plane components of $\mathbf{e}$ emerge reversed and the
azimuth advances by exactly
\begin{equation}
\Omega_{e,n+1}=\Omega_{e,n}+\pi
\label{eq:halfturn}
\end{equation}
per cycle (at $j_z=0$, $\epsh=0$), a geometric statement independent
of $C_K$. On a rotating cycle $\mathbf{e}$ never crosses
$\hat{\mathbf{z}}$, so the azimuth at $j_z=0$ is constant, and a
finite $j_z$ adds only a slow continuous precession at the quadrupole
rate $\left\langle f_\Omega\right\rangle j_z$, where
$\left\langle\cdot\right\rangle$ denotes, here and throughout, the time
average over one cycle. This average of Equation \ref{eq:fOmega} has
the closed form $\left\langle f_\Omega\right\rangle
=\left(6E-3K\right)/\left(4K\right)$ evaluated at $1/k$ (\citekdm),
where $E$ is the complete elliptic integral of the second kind, also
as a function of its parameter.
Multiplied by the cycle period, the precession rate gives the advance
of $\Omega_e$ per cycle, $F_\text{rot}\,j_z$, with the rotating phase slip
\begin{equation}
F_\text{rot}=T_e\left\langle f_\Omega\right\rangle
=\frac{3\left[2E\left(1/k\right)-K\left(1/k\right)\right]}
{2\sqrt{C_x}},
\label{eq:Frot}
\end{equation}
which vanishes at $C_K\simeq0.112$, the zero of
$\left\langle f_\Omega\right\rangle$ found by \citekdm. On the
librating side a finite $j_z$ adds, on top of the half-turn of
Equation \ref{eq:halfturn}, a slow precession that advances $\Omega_e$
by $F_\text{lib}\,j_z$ per cycle, with the librating phase slip
\begin{equation}
F_\text{lib}\left(C_K\right)=\frac{3}{2k\sqrt{C_z}}\left[2E\left(k\right)
-\left(2-k\right)K\left(k\right)\right]<0,
\label{eq:F}
\end{equation}
the librating counterpart of $F_\text{rot}$ ($F$ in Equation 16 of
\citepaperIV).

\subsection{Instantaneous torque and per-cycle kicks}
\label{sec:kick}

Equation \ref{eq:milankovitch_j} gives
$\dot j_z=\partial\phi/\partial\varphi_z$, the derivative of $\phi$ with
respect to the angle $\varphi_z$ of a rigid rotation of $\mathbf{e}$ and
$\mathbf{j}$ together about $\hat{\mathbf{z}}$. The axisymmetric $f_0$,
unchanged by such rotations, therefore does not change $j_z$ at any
instant. The $f_2$ term, in contrast, depends on $\varphi_z$ only through
$\cos2\varphi_z$ and $\sin2\varphi_z$ (Section \ref{sec:setup}), so at
every state
\begin{equation}
\dot j_z^{\rm hex}=\epstwo\left(A_2\sin2\Omega_e+B_2\cos2\Omega_e\right),
\label{eq:djzhex}
\end{equation}
where $A_2$ and $B_2$ depend only on the rotation invariants $e$,
$i_e$, $j_z$ and $s\equiv\left(\mathbf{e}\times\mathbf{j}\right)_z$
(explicit terms given in Equation \ref{eq:A2B2}), and $\Omega_e$ factors
out of every
cycle integral below. Since $B_2\propto j_z$ at every state, it
vanishes on the exact cycles of Section \ref{sec:cycles} ($j_z=0$),
where $f_2=f_2\left(\Omega_e{=}0\right)\cos2\Omega_e$, with
$f_2\left(\Omega_e{=}0\right)$ taken at the same point of the cycle, a
quadratic polynomial in $e_x^2$, $e_z^2$ and $j_y^2$ (explicit terms
given in Equation \ref{eq:f2zero}). The instantaneous torque is
therefore $\dot j_z=-2\epstwo\,f_2\left(\Omega_e{=}0\right)\sin2\Omega_e$, an exact
second harmonic before any averaging.

Since $f_2$ is even in $\mathbf{e}$ and
$\sin2\Omega_e$ is invariant under the half-turn
(Equation \ref{eq:halfturn}), the cycle average at frozen azimuth
survives on both classes and only replaces $f_2\left(\Omega_e{=}0\right)$ by its
integral over the cycle. The per-cycle kick, the hexadecapole analogue of
the $\left\langle f_j\right\rangle$ of \citekdm\ quoted per cycle rather
than per unit time, is given by
\begin{equation}
\begin{gathered}
\Delta j_z=-\epstwo\,\chi\left(C_K\right)\sin2\Omega_e ,\\
\chi=2\int_0^{T_e}f_2\left(\Omega_e{=}0\right)\,\mathrm{d}\tau ,
\end{gathered}
\label{eq:kick}
\end{equation}
where the integral runs over one KLC at $j_z=0$, of either class.
These are the simple pendulums of Section \ref{sec:cycles}: along the
cycle $e_x$, $e_z$ and $j_y$ are Jacobi elliptic functions of $\tau$, of
parameter $k$ or $1/k$, so the integral of the polynomial
$f_2\left(\Omega_e{=}0\right)$ over one cycle reduces to complete
elliptic integrals, giving
\begin{align}
\chiH^\text{lib}&=\frac{\sqrt6}{\sqrt{1-C_K}}\Bigg[
\frac{\left(1-C_K\right)\left(1186\,C_K-1761\right)}{1440}
E\left(k\right)\nonumber\\
&\qquad+\frac{1352\,C_K^2-362\,C_K-45}{576}K\left(k\right)\Bigg]
\label{eq:chiH}
\end{align}
on the librating class and
\begin{align}
\chiH^\text{rot}&=\sqrt2\,\sqrt{2C_K+3}\Bigg[
\frac{1186\,C_K-1761}{1440}E\left(1/k\right)\nonumber\\
&\qquad+\frac{842\,C_K-45}{576}K\left(1/k\right)\Bigg]
\label{eq:chiHrot}
\end{align}
on the rotating class.

\begin{figure}
 \begin{centering}
 \includegraphics[width=0.88\columnwidth]{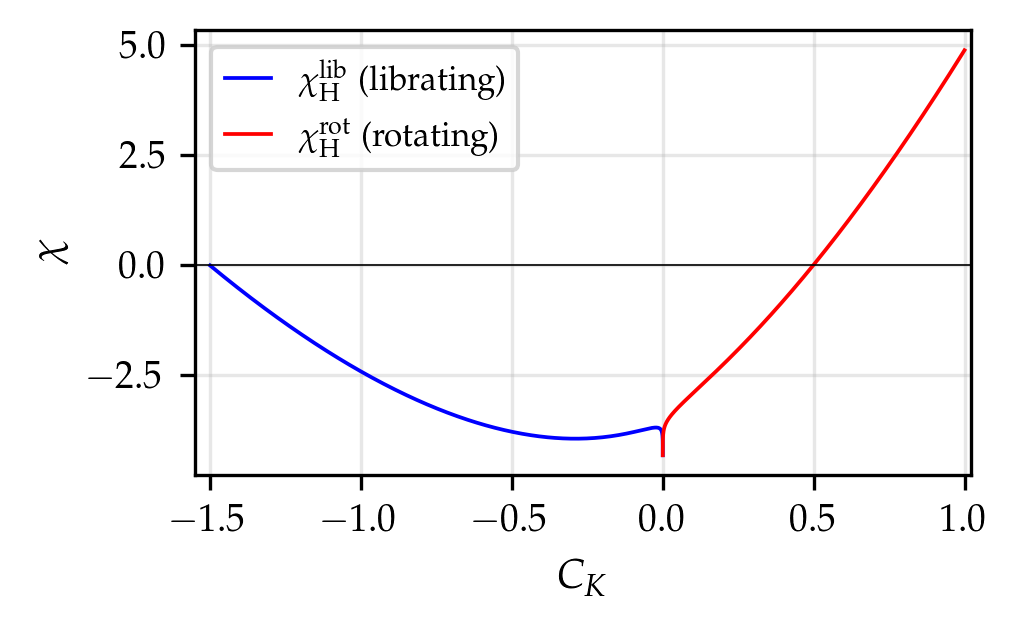}
 \par\end{centering}
 \caption{The closed-form per-cycle kicks of the hexadecapole on the
 two KLC classes, $\chiH^\text{lib}$ on the librating class and
 $\chiH^\text{rot}$ on the rotating class (Equations
 \ref{eq:chiH}--\ref{eq:chiHrot}).
 \label{fig:chi}}
\end{figure}

The kicks of both classes are shown in Figure \ref{fig:chi}. On the
librating class $\chiH^\text{lib}<0$ across the entire window. Its magnitude
vanishes linearly at the deep edge $C_K\rightarrow-\frac{3}{2}$, has a
local maximum $\left|\chiH^\text{lib}\right|\simeq3.94$ at $C_K\simeq-0.29$, and
diverges as $C_K\rightarrow0^-$; the
librating drive never vanishes. The rotating kick, by contrast,
changes sign at the root of Equation \ref{eq:chiHrot}, $C_K\simeq0.4984$,
the hexadecapole analogue of the zero of the rotating octupole drive
at $C_K=\frac{4}{11}$ (\citekdm). Both kicks diverge at $C_K=0$ only
through the cycle period, which diverges logarithmically there: the
kick per unit time, $\chi/T_e$, stays finite on both classes.

\subsection{Rotating cycles}
\label{sec:rotclass}

On the rotating class the azimuth, $\Omega_e$, is a smooth slow variable
(Section \ref{sec:cycles}), so the slow system, now for the hexadecapole,
takes the rate form of \citekdm\ and \citepaperI,
\begin{equation}
\frac{\mathrm{d}\Omega_e}{\mathrm{d}\tau}
=\left\langle f_\Omega\right\rangle j_z,\qquad
\frac{\mathrm{d}j_z}{\mathrm{d}\tau}
=-\epstwo\,\frac{\chiH^\text{rot}}{T_e}\sin2\Omega_e ,
\label{eq:rotrates}
\end{equation}
with the hexadecapole drive in the role of the octupole term
$\left\langle f_j\right\rangle$ of \citekdm. Taking $C_K=C_K^0$, its
initial value, as in Papers I--IV, Equations \ref{eq:rotrates} are the
equations of a simple pendulum in the doubled angle $2\Omega_e$. The
two sign changes ($F_\text{rot}$
at $C_K\simeq0.112$, $\chiH^\text{rot}$ at $C_K\simeq0.4984$) divide the rotating
window into three regimes: the stable azimuths are $\Omega_e=0,\pi$
below $C_K\simeq0.112$ and above $0.4984$, and $\Omega_e=\pm\pi/2$ between
them.

\subsection{Librating cycles}
\label{sec:pendulum}

On the librating class the azimuth changes by $\pi$ at each passage
through the eccentricity minimum $e_{\min}$ of the quadrupole KLC, in the
$j_z=0$ limit (Equation \ref{eq:halfturn}). To leading order in $\epstwo$ and $j_z$ the per-cycle
dynamics can be represented by the staggered map
\begin{equation}
\begin{gathered}
\Omega_{e,n+1}=\Omega_{e,n}+\pi+F_\text{lib}\,\Jh_n,\\
\Jh_{n+1}=\Jh_n-\epstwo\,\chiH^\text{lib}\sin2\Omega_{e,n+1},
\end{gathered}
\label{eq:map}
\end{equation}
where $\Omega_{e,n}$ is the azimuth at the $n$-th $e_{\max}$ passage and
the slow momentum $\Jh_n$ is the value of $j_z$ at the $e_{\min}$ passage
that follows it, halfway between successive azimuth registrations; for
an initial condition at the quadrupole $e_{\min}$, $\Jh_0=j_{z,0}$. Because $f_2$ is
even in $\mathbf{e}$, the kick term of Equation \ref{eq:map} does not alternate, and unlike the octupole
kicks of \citepaperIV\ it leaves no alternating component to remove
from $j_z$ (compare with Equation 17 of \citepaperIV). With the
parity-corrected phase $\psi=\Omega_e+n\pi$
(modulo $2\pi$) the continuum limit is directly
\begin{equation}
\begin{gathered}
\frac{\mathrm{d}\psi}{\mathrm{d}n}=F_\text{lib}\Jh,\quad
\frac{\mathrm{d}\Jh}{\mathrm{d}n}=-\epstwo\,\chiH^\text{lib}\sin2\psi,\\
H=\frac{F_\text{lib}}{2}\Jh^2-\frac{\epstwo\chiH^\text{lib}}{2}\cos2\psi.
\end{gathered}
\label{eq:pendulum}
\end{equation}
Taking $C_K=C_K^0$ as in Section \ref{sec:rotclass}, Equations
\ref{eq:pendulum} are the equations of a simple pendulum in the doubled
angle $2\psi$, with $\Jh$ as the momentum and a torque of first order
in $\epstwo$. Since $\chiH^\text{lib}<0$ as well as $F_\text{lib}$, the stable
fixed points are at $\psi=0,\pi$, the \textit{aligned} configurations,
in which the projection of $\mathbf{e}$ on the outer orbital plane lies
along the outer apsidal line. The librating octupole pendulum of
\citepaperIV\ has the same structure, in the same doubled angle, with
two differences: its torque is of second order in $\epsilon_\text{oct}$,
and it is stable at $\psi=\pm\pi/2$, where the hexadecapole pendulum
has its saddles.

\begin{figure}
 \begin{centering}
 \includegraphics[width=0.92\columnwidth]{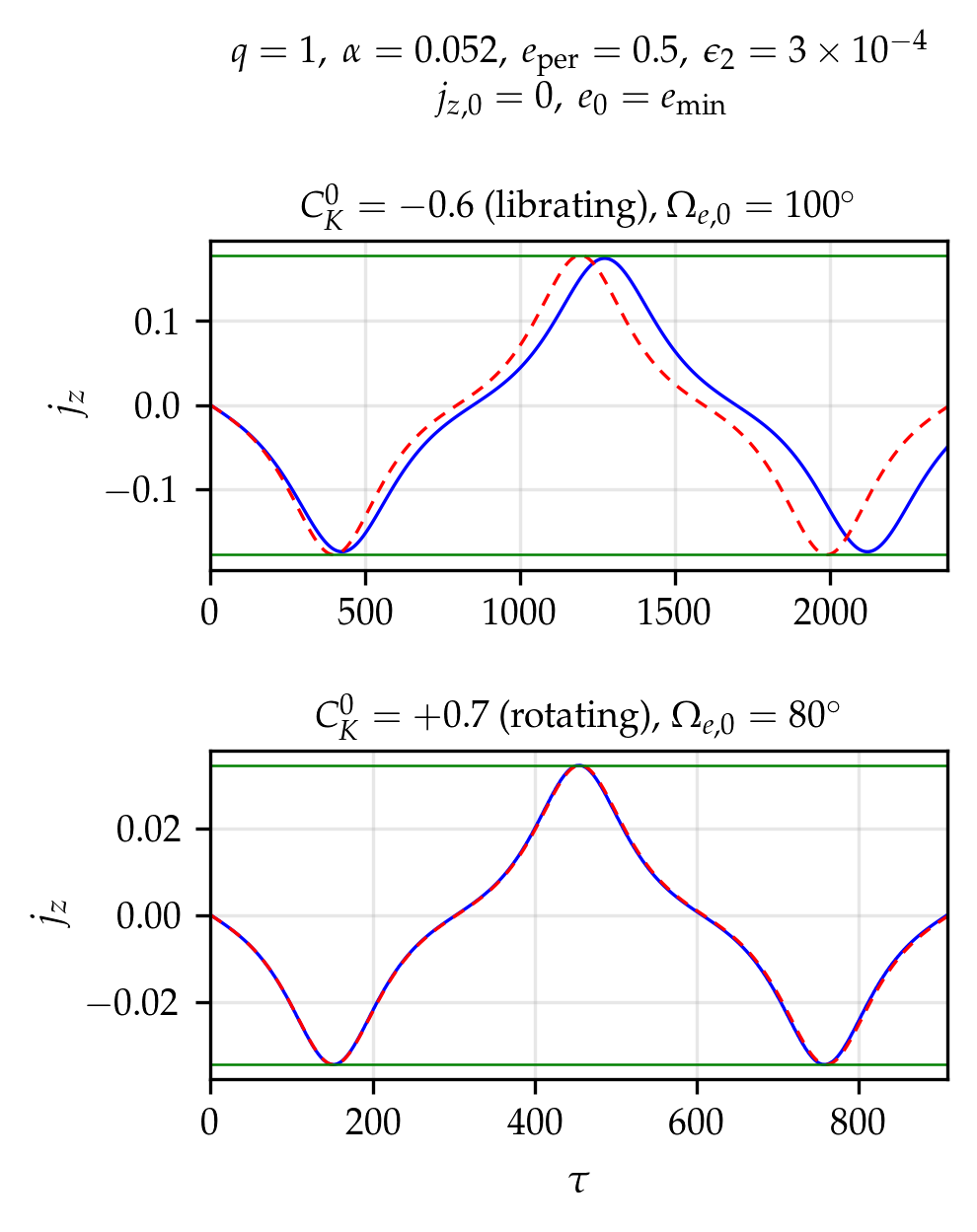}
 \par\end{centering}
 \caption{Results of a numerical integration of the DA equations (blue)
 and a numerical solution of the doubled-angle pendulum (Equations
 \ref{eq:rotrates} and \ref{eq:pendulum}; red dashed), whose
 coefficients and initial state are taken from the initial
 conditions, one example per class, with the parameters above the
 panels. Top panel: $j_z$ versus the
 normalized time $\tau$ on the librating class. Bottom panel: the same
 on the rotating class. The green lines mark the predicted extrema
 $\pm j_z^{\max}$ (Equation \ref{eq:jmax}).
 \label{fig:evolution}}
\end{figure}

\needspace{4\baselineskip}
\subsection{Maximal deviation of \texorpdfstring{$j_z$}{jz} from \texorpdfstring{$j_{z,0}=0$}{jz0=0}}
\label{sec:landscape}

Without a class label, $F$ and $\chi$ denote the pair of either class,
$\left(F_\text{lib},\chiH^\text{lib}\right)$ or
$\left(F_\text{rot},\chiH^\text{rot}\right)$, and on the rotating class
$\psi$ is $\Omega_e$ itself. Energy conservation solves the pendulum of
both classes. Since $j_z$ is its velocity,
$\mathrm{d}\psi/\mathrm{d}n=F\,j_z$, an initial condition
$\left(\psi_0,\,j_z=0\right)$ starts it at rest, so that off the four
equilibrium azimuths, $\psi=0,\pm\pi/2,\pi$, it librates about the
nearest stable azimuth $\psi_s$, which is $0$ or $\pi$ where $\chi/F>0$
and $\pm\pi/2$ where $\chi/F<0$ (Sections \ref{sec:rotclass} and
\ref{sec:pendulum}), and
\begin{equation}
\begin{gathered}
j_z^2\left(\psi\right)=\frac{\epstwo\chi}{F}
\left(\cos2\psi-\cos2\psi_0\right),\\
j_z^{\max}=J_\text{env}\left|\sin\left(\psi_0-\psi_s\right)\right| ,\quad
J_\text{env}\equiv\sqrt{2\epstwo\left|\chi/F\right|} ,
\end{gathered}
\label{eq:jmax}
\end{equation}
where $J_\text{env}$ is the envelope over initial azimuths. An example of a numerical integration of the DA equations
compared with that of the simple-pendulum equations (Equations
\ref{eq:rotrates} and \ref{eq:pendulum}) is shown in
Figure \ref{fig:evolution} for each class. As can be seen, the
pendulum follows the DA integration on both classes, up to a phase
drift that accumulates at finite $\epstwo$. The drift comes from terms
that the frozen pendulum omits, of relative order $\epstwo/\left|F\right|$:
the advance of $\Omega_e$ by the axisymmetric part $f_0$ of the
hexadecapole; the dependence of $\chi$ on $j_z$, which through the
pendulum Hamiltonian (Equation \ref{eq:pendulum}) adds the precession
$-\frac{\epstwo}{2}\left(\mathrm{d}\chi/\mathrm{d}j_z\right)\cos2\psi$; and
the drift of $C_K$ with $j_z$ (\citepaperIV). Being of first order in $\epstwo$, the excursions scale as
$\sqrt{\epstwo}$ and the slow periods as $\epstwo^{-1/2}$ cycles, as for
the rotating octupole pendulum of \citepaperI, whereas the librating
octupole pendulum of \citepaperIV, of second order, scales as
$\epsilon_\text{oct}$ and $\epsilon_\text{oct}^{-1}$.

\begin{figure}
 \begin{centering}
 \includegraphics[width=0.92\columnwidth]{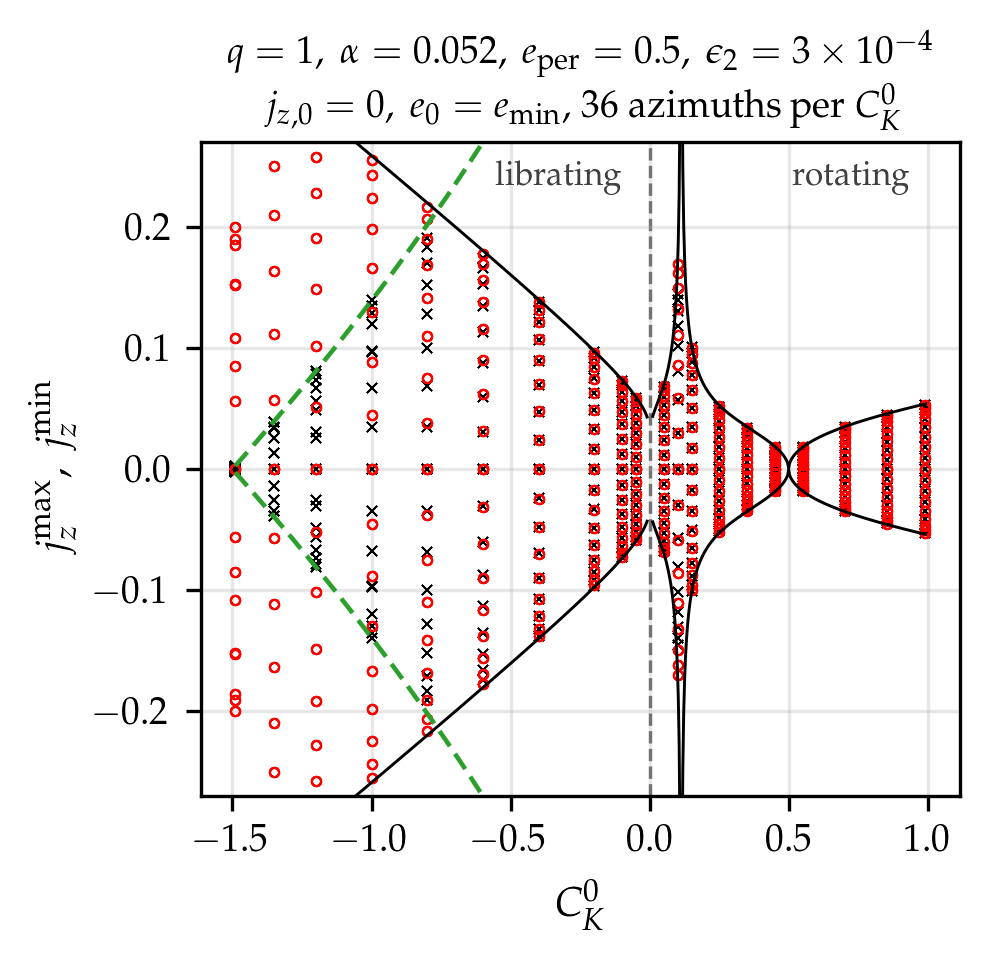}
 \par\end{centering}
 \caption{The extremes $j_z^{\max}$ and $j_z^{\min}$ versus $C_K^0$ for the
 integrations of Section \ref{sec:landscape}, with the parameters above
 the panel. The measured extrema are shown with black crosses, the
 frozen-pendulum prediction for each initial condition with open red
 circles, the envelope $\pm J_\text{env}$ (Equation \ref{eq:jmax}) with
 black lines, and the fixed-point ceiling $\pm J_\text{ceil}$
 (Equation \ref{eq:Jceil}) with green dashed lines.
 \label{fig:jzminmax}}
\end{figure}

The maximal and minimal values of $j_z$ starting from $j_{z,0}=0$,
obtained in $720$ numerical integrations of the full DA equations
(quadrupole and hexadecapole) at $q=1$, $\eper=0.5$ and $\alpha=0.052$,
are compared with the prediction of Equation \ref{eq:jmax} in
Figure \ref{fig:jzminmax}. The initial conditions are at the
quadrupole $e_{\min}$ \citep[known
analytically;][]{vashkovyak1999,kinoshita2007,antognini2015,basha2025},
with $36$ azimuths for each of twenty values of
$C_K^0$ from $-1.49$ to $+0.99$, and the coefficients are evaluated at
them. Each integration covers up to three periods of small oscillation
of the frozen pendulum, $2\pi/\sqrt{2\epstwo\left|\chi F\right|}$
KLCs. The
excursions at each initial azimuth are shown in
Appendix \ref{sec:profiles}, also for $\alpha=0.026$. The
profile of Equation \ref{eq:jmax} agrees with the numerical results
except in two places. Near the zero of $F_\text{rot}$
($C_K\simeq0.112$) the envelope diverges, and the approximation is not
uniform there. Deep in the librating window the frozen prediction
overshoots, and the measurements are capped instead by the fixed-point
ceiling of \citepaperIV\ (Equation 28 therein, with $j_{z,0}=0$),
\begin{equation}
J_\text{ceil}=\frac{1}{4}\left(\sqrt{15}-\sqrt{15-8\left(C_K^0+\frac{3}{2}\right)}\right).
\label{eq:Jceil}
\end{equation}

To leading order, $\alpha$ and $\eper$ enter only through $\epstwo$: the
maximal excursion follows $\sqrt{\epstwo}\propto\alpha$, and at fixed $\alpha$ the azimuth profiles
for different $\eper$ collapse under the $\sqrt{\epstwo}$ scaling onto one
profile per value of $C_K^0$ (Figure \ref{fig:scalings} in
Appendix \ref{sec:profiles}). The ceiling $J_\text{ceil}$, in contrast,
does not depend on $\alpha$. The remaining departures at finite
$\epstwo$ come from the variation of $C_K$ and the corrections to the
phase slip, which the frozen pendulum does not describe. The leading
excursion of $j_z$ is thus given in closed form on both KLC classes.

\begin{figure}
 \begin{centering}
 \includegraphics[width=0.92\columnwidth]{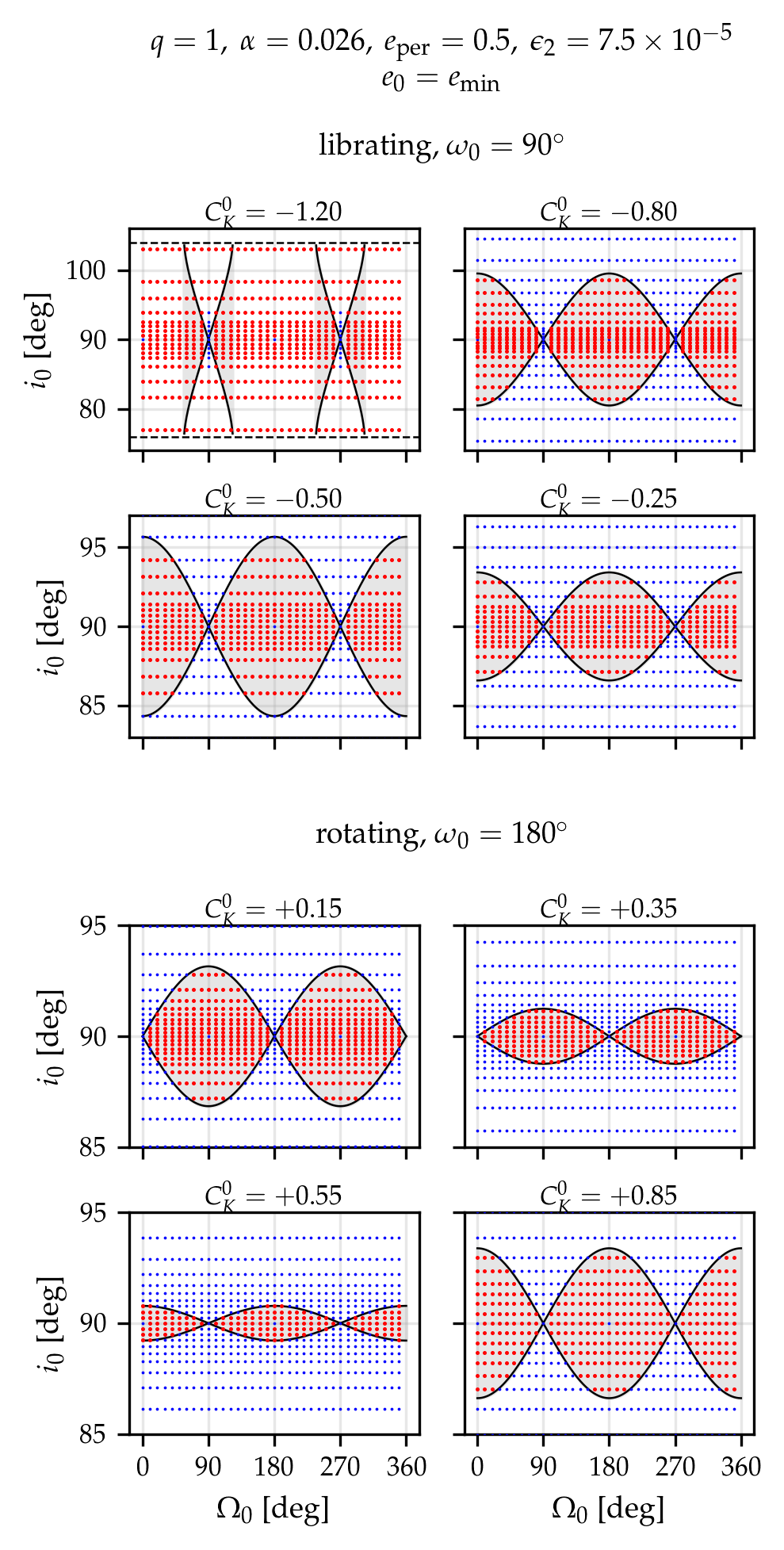}
 \par\end{centering}
 \caption{The hexadecapole flip lens in the observable
 $\left(\Omega_0,i_0\right)$ plane, one value of $C_K^0$ per panel, on
 the librating class (top four panels) and the rotating class (bottom
 four), with the parameters above the panels. Red points mark
 integrations in which $j_z$ crossed zero and blue points those in
 which $j_z$ kept its sign. The gray lens
 bounded by the black curves is the flip criterion of
 Section \ref{sec:flips}. At $C_K^0=-1.2$ the dashed lines bound the
 band in which librating cycles exist.
 \label{fig:flipplane}}
\end{figure}

\subsection{Flip criterion}
\label{sec:flips}

An orbit flips when $j_z$ crosses zero. Since $j_z$ is the velocity of
the doubled-angle pendulum, this happens exactly when the pendulum
librates,
$\left|j_{z,0}\right|<J_\text{env}\left|\cos\left(\psi_0-\psi_s\right)\right|$,
with $J_\text{env}$ and $\psi_s$ as in Equation \ref{eq:jmax}: by
the energy relation the excursion from $j_z=0$ is largest at the saddle
azimuths, while the flip window is widest around the stable ones. The
boundary agrees with numerical flip maps on both classes.
Figure \ref{fig:flipplane} shows it in the plane of two observable
orbital elements, the initial longitude of the ascending node
$\Omega_0$ (measured, like $\Omega_e$, from $\hat{\mathbf{x}}$) and
the initial inclination $i_0$, for initial conditions at the quadrupole
$e_{\min}$, librating in its top four panels and rotating in the bottom
four. Since $\mathbf{e}$ moves in the orbital plane,
which at $j_z=0$ is spanned by $\hat{\mathbf{z}}$ and the line of nodes,
$\psi_0=\Omega_0$ modulo $\pi$ (as in \citepaperIV), the azimuths
registered at the $e_{\max}$ passages before and after the initial
condition lying at $\Omega_0\mp F_\text{lib}\,j_{z,0}/2$. The resulting hexadecapole
flip lens has its pinches at $\Omega_0=90^\circ,270^\circ$ and exactly
symmetric lobes, since at $q=1$ there is no odd-multipole kick offset;
the octupole lens of \citepaperIV\ has its pinches at
$\Omega_0=0^\circ,180^\circ$ and lobes made asymmetric by that offset.
At $C_K^0=-1.2$ the predicted lens covers most of the band in which
librating cycles exist, which the fixed-point floor of \citepaperIV\
bounds, and the integrations with $j_{z,0}\neq0$ flip throughout the
band except at the saddle azimuths, where the lens pinches. The saddle
starts keep the sign of $j_z$ up to $\left|j_{z,0}\right|=0.03$, but from
$\left|j_{z,0}\right|=0.045$ they flip, although the criterion forbids
it. At this depth $F_\text{lib}$ is small, and the precession from the
$j_z$-dependence of $\chi$ cancels it on the saddle azimuths at finite
$j_z$, where the slow dynamics acquires two stable fixed points, one for
each sign of $j_z$; the saddle starts with small $\left|j_{z,0}\right|$
librate about them, keeping their sign. On the rotating class the lens exists as well. Its orientation follows the stable
azimuths: between the phase-slip zero ($C_K\simeq0.112$) and the drive zero
($C_K\simeq0.4984$) the pinches are at $\Omega_0=0^\circ,180^\circ$, and
above $0.4984$ at $90^\circ,270^\circ$, as on the librating class.

\needspace{5\baselineskip}
\section{Unequal-mass inner binaries}
\label{sec:competition}

For $q<1$ the octupole no longer vanishes,
$\epsilon_\text{oct}\propto\left(1-q\right)/\left(1+q\right)$, and the
two KLC classes differ in the order at which it competes with the
hexadecapole.

\needspace{4\baselineskip}
\subsection{Rotating cycles}
\label{sec:rotoct}

On the rotating class the octupole is a first-order rate,
$O\left(\epsilon_\text{oct}\right)$, while the hexadecapole drive is
$O\left(\epstwo\right)$, with $\epsilon_\text{oct}\propto\alpha$ and
$\epstwo\propto\alpha^2$ (Equation \ref{eq:strengths}). For any $q<1$ the octupole therefore
dominates as $\alpha\to0$, except near the zero of its drive at
$C_K=\frac{4}{11}$ (\citekdm). Since
$\epstwo\left(q{=}1\right)=\frac{1}{4}\epsilon^2_\text{oct}
\left(q{=}0\right)$, away from that zero the two drives are comparable only
for nearly equal masses,
$1-q\lesssim\epsilon_\text{oct}\left(q{=}0\right)$, that is
$q\gtrsim0.99$ for a fiducial
$\left(\alpha,\eper\right)=\left(0.026,0.5\right)$. The phase slip
$F_\text{rot}$ (Equation \ref{eq:Frot}) is common to the octupole and
hexadecapole pendulums and vanishes at $C_K\simeq0.112$, where the leading
pendulum reduction fails.

\subsection{Librating cycles}
\label{sec:liboct}

On the librating class the half-turn (Equation \ref{eq:halfturn})
survives the axisymmetric part $f_0$ of the hexadecapole: like the quadrupole, $f_0$ is symmetric
under reflection through the \textit{x-z} plane of
Section \ref{sec:cycles}, so the $j_z=0$ cycle keeps $j_x=j_z=e_y=0$.
A per-cycle kick built from a term \textit{odd} in
$\mathbf{e}$ therefore alternates and cancels pairwise, leaving the
octupole slow torque at second order,
$-\frac{\epsilon^2_\text{oct}\Lambda}{4}\sin2\psi$ (Equation 18 of
\citepaperIV), where $\Lambda\left(C_K\right)>0$ is the slow-pendulum
torque coefficient (Equation 19 therein). The
kick of the even hexadecapole does not alternate
(Section \ref{sec:pendulum}). Under the librating half-turn, odd
multipoles act on the slow dynamics only at second order in their
strength, and even multipoles at first order. The hexadecapole thus
enters at the same order in $\alpha$ as the octupole slow torque,
$O\left(\alpha^2\right)$ (Section \ref{sec:setup}), and in the same
harmonic, so that the slow momentum evolves as
\begin{equation}
\begin{gathered}
\frac{\mathrm{d}\Jh}{\mathrm{d}n}=-C_\text{tot}\sin2\psi,\\
C_\text{tot}=\frac{\epsilon^2_\text{oct}\Lambda}{4}+\epstwo\chiH^\text{lib} .
\end{gathered}
\label{eq:lameff}
\end{equation}
Here $\psi$ is the same parity-corrected phase as in
Section \ref{sec:pendulum},
still with $\mathrm{d}\psi/\mathrm{d}n=F_\text{lib}\Jh$, and $\Jh$ is
the slow momentum of \citepaperIV\ (Equation 17 therein): $j_z$ at the
$e_{\min}$ passages with the zigzag of the alternating octupole kicks
removed, $j_z=\Jh\pm b\sin\psi$ with $b=\epsilon_\text{oct}\chi_\text{oct}/2$
and $\chi_\text{oct}$ the octupole per-cycle kick ($\chi$ in Equation 13
of \citepaperIV). At $q=1$ it is the $\Jh$ of Section \ref{sec:pendulum},
and for an initial condition at the quadrupole $e_{\min}$ with
$\omega_0=90^\circ$ it starts from
$\Jh_0=j_{z,0}-b\sin\Omega_0$, with $\Omega_0$ the initial node
(\citepaperIV). The
torque is kept at $O\left(\alpha^2\right)$; terms of higher order in
$\alpha$ are dropped.

\begin{figure}
 \begin{centering}
 \includegraphics[width=0.92\columnwidth]{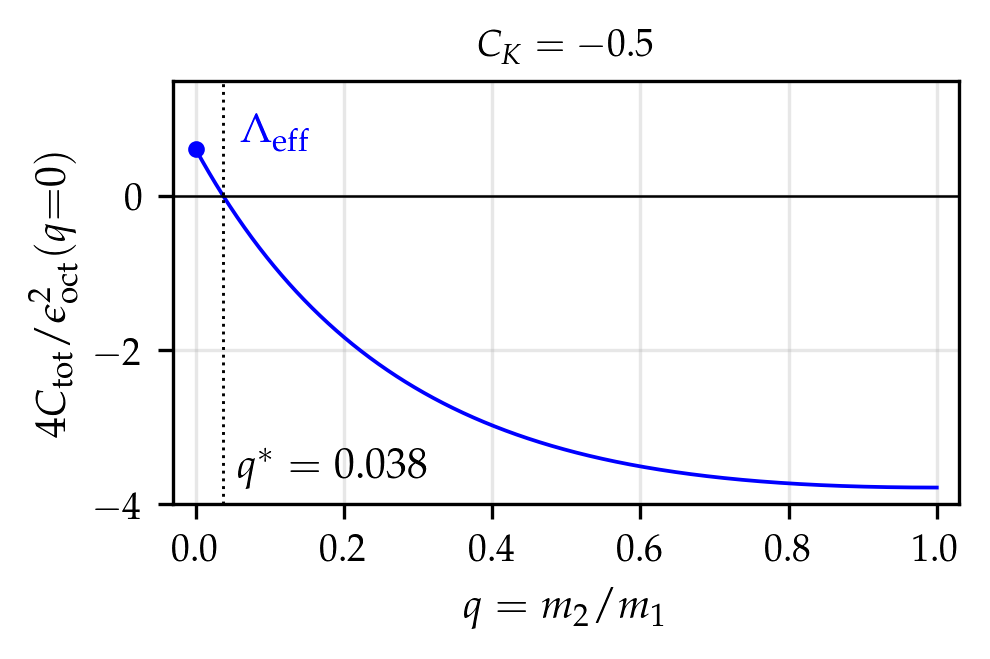}
 \par\end{centering}
 \caption{The combined librating torque coefficient of
 Equation \ref{eq:lameff} in units of
 $\epsilon^2_\text{oct}\left(q{=}0\right)/4$
 (Equation \ref{eq:ctotnorm}) versus $q$, for an example value of $C_K$
 given above the panel.
 It equals $\Lambda_\text{eff}$ at $q=0$ and vanishes at
 $q^{*}\simeq0.038$ (dotted).
 \label{fig:ctotq}}
\end{figure}

The two torques have opposite signs ($\Lambda>0>\chiH^\text{lib}$),
and the ratio of their magnitudes,
$\frac{4\left|\chiH^\text{lib}\right|}{\Lambda}\frac{\epstwo}{\epsilon^2_\text{oct}}
=\frac{4\left|\chiH^\text{lib}\right|}{\Lambda}\frac{1+q^3}{\left(1+q\right)\left(1-q\right)^2}$
(Section \ref{sec:setup}), grows with $q$, and $C_\text{tot}$ vanishes
where this ratio is unity. At $q=0$, where $\epstwo=\epsilon^2_\text{oct}$, the
ratio is $4\left|\chiH^\text{lib}\right|/\Lambda$. It rises from exactly
$16/17$ at the deep edge of the librating window and exceeds $1.2$ only
within about $0.002$ of the separatrix, where it diverges: the two
torques nearly cancel, and the combined torque takes the form of the
octupole slow torque with $\Lambda$ replaced by
$\Lambda_\text{eff}\equiv\Lambda+4\chiH^\text{lib}$,
$C_\text{tot}=\epsilon^2_\text{oct}\Lambda_\text{eff}/4$.
$\Lambda_\text{eff}$ changes sign, and the ratio crosses unity, at
\begin{equation}
C_K^{*}\simeq-0.146 .
\label{eq:ckstar}
\end{equation}
Above $C_K^{*}$ ($C_K^{*}<C_K<0$) the combined torque has the
hexadecapole sign for all $q$, and the stable azimuths of the combined
slow pendulum are the aligned ones, the saddles of the octupole-only
theory. Below $C_K^{*}$ ($C_K<C_K^{*}$) it changes sign at a mass ratio
$q^{*}$. Since $\epsilon^2_\text{oct}/\epstwo\simeq1-q$ for small $q$,
$q^{*}\simeq\Lambda_\text{eff}/\Lambda$, the fraction of the octupole
torque left uncancelled at $q=0$, which is below $0.06$ for every
$C_K<C_K^{*}$; $q^{*}\simeq0.038$ at $C_K=-0.5$. In units of
$\epsilon^2_\text{oct}\left(q{=}0\right)/4$ the combined coefficient,
\begin{equation}
\frac{4C_\text{tot}}{\epsilon^2_\text{oct}\left(q{=}0\right)}
=\Lambda\left(\frac{1-q}{1+q}\right)^2
+4\chiH^\text{lib}\,\frac{1+q^3}{\left(1+q\right)^3},
\label{eq:ctotnorm}
\end{equation}
depends on $q$ and $C_K$ alone, so that $q^{*}$, at
$O\left(\alpha^2\right)$, is independent of $\alpha$ and $\eper$; it
depends on $C_K$ and approaches zero as $C_K\to C_K^{*}$ from below.
Figure \ref{fig:ctotq} shows this coefficient versus $q$ at
$C_K=-0.5$: it equals $\Lambda_\text{eff}$ at $q=0$ and changes sign at
$q^{*}$.

\begin{figure}
 \begin{centering}
 \includegraphics[width=0.92\columnwidth]{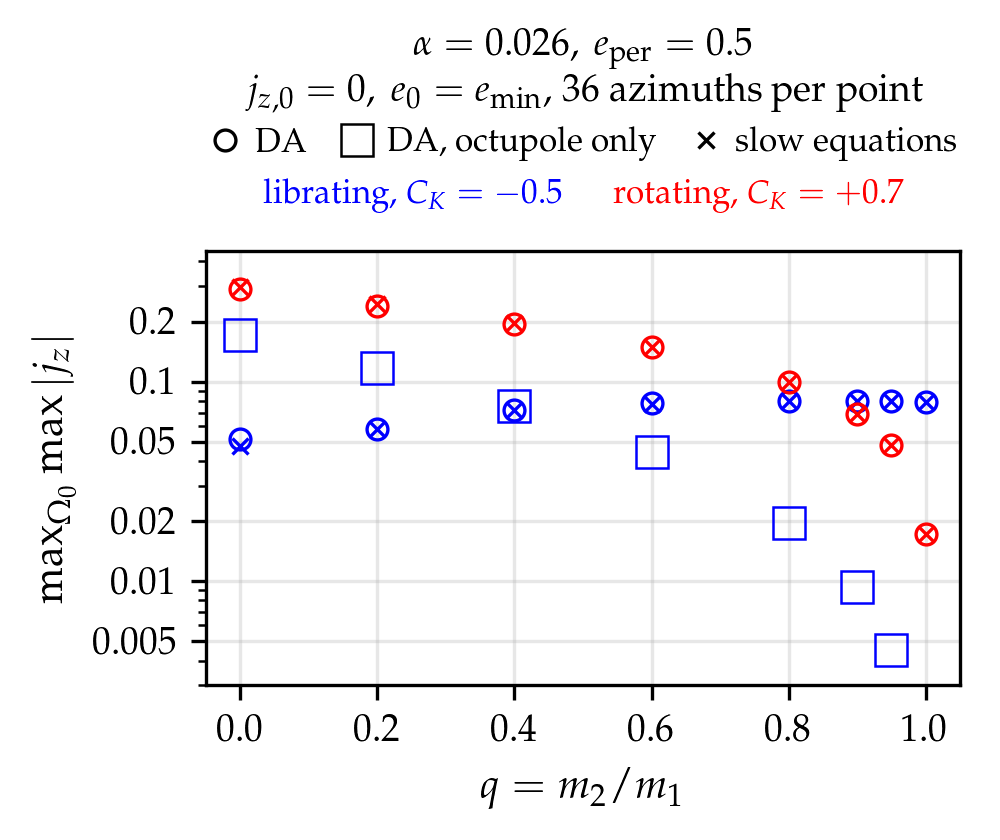}
 \par\end{centering}
 \caption{The largest excursion of $j_z$ versus the inner mass ratio,
 with the parameters and initial conditions above the panel:
 $\max\left|j_z\right|$ over initial azimuths on the librating
 ($C_K=-0.5$, blue) and rotating ($C_K=+0.7$, red) classes, from
 numerical integrations of the DA equations (circles), on the librating
 class also truncated at the octupole (squares), and from the slow
 equations (crosses): on the librating class Equation \ref{eq:lameff}
 from the initial slow momentum $\Jh_0=-b\sin\Omega_0$ of
 Section \ref{sec:liboct} ($j_{z,0}=0$), with the zigzag
 $j_z=\Jh\pm b\sin\psi$, and on the rotating class
 Equation \ref{eq:rotrates} with the octupole term
 $\left\langle f_j\right\rangle$ of \citekdm\ included.
 \label{fig:qsweep}}
\end{figure}

Figure \ref{fig:qsweep} shows the largest excursion of $j_z$ over
initial azimuths versus $q$ for examples from the two classes, in
numerical integrations of the DA equations and from the slow equations.
From $q=1$ to $q=0.95$ the excursion grows about threefold
on the rotating class, where the octupole acts at first order, but
barely changes on the librating class, where it acts at second order.
Truncated at the octupole (squares), the librating excursion instead
falls nearly in proportion to $\epsilon_\text{oct}$, to $0.0045$ at
$q=0.95$, and vanishes at $q=1$; on the rotating class the truncation
changes it by less than $0.2$ per cent (not shown).

\begin{figure}
 \begin{centering}
 \includegraphics[width=0.92\columnwidth]{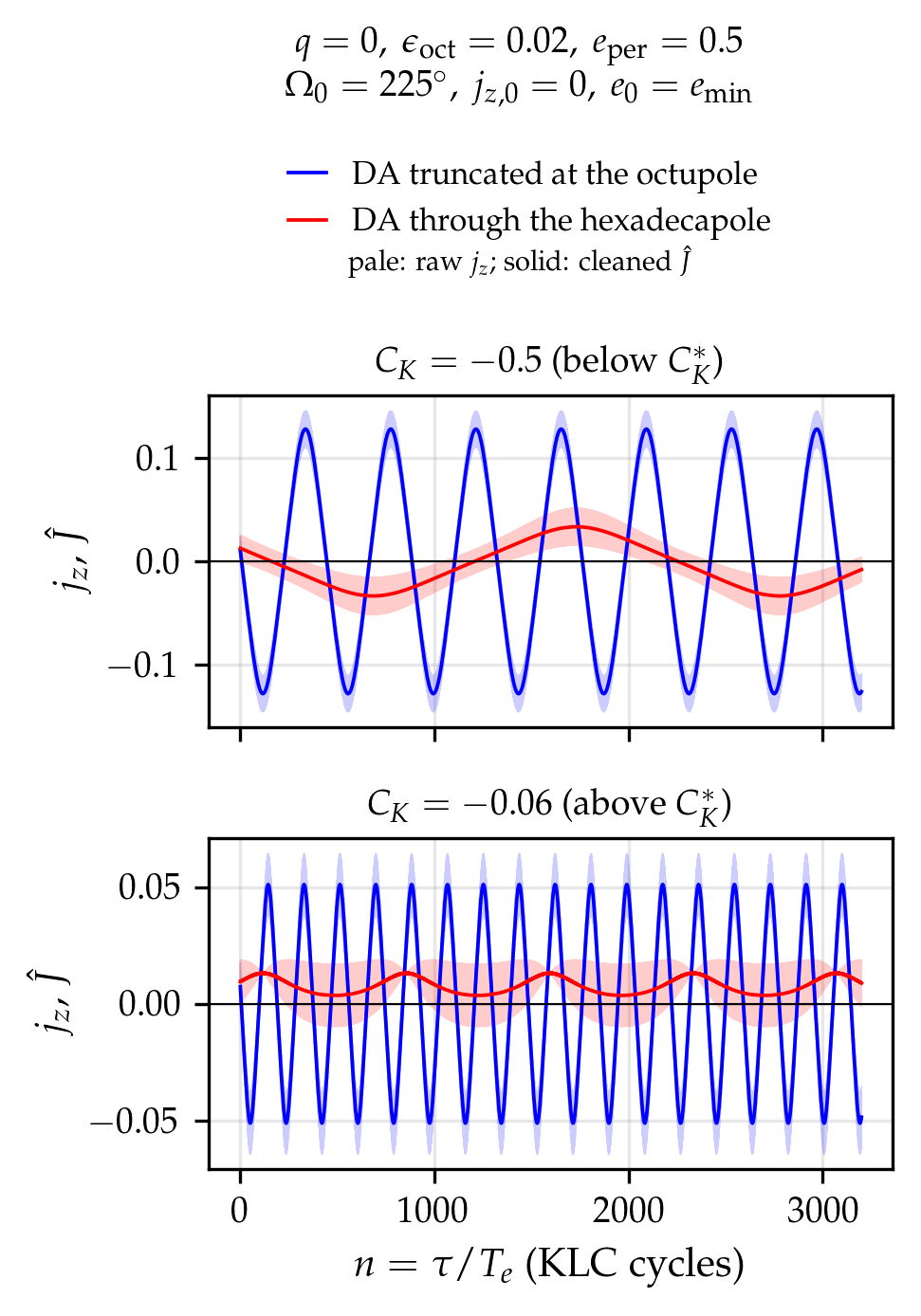}
 \par\end{centering}
 \caption{Numerical integrations of the DA equations truncated at the
 octupole (blue) and through the hexadecapole (red), from the same
 initial conditions, with the shared parameters and initial conditions
 above the panels and $C_K$ above each panel. Pale curves show the raw
 $j_z$ and solid curves the cleaned slow momentum $\Jh$, versus the
 number of KLCs $n$.
 \label{fig:octtrunc}}
\end{figure}

Through the near cancellation, the hexadecapole changes the leading
librating slow torque by $O\left(1\right)$ in the limit $q\to0$, and above
$C_K^{*}$ reverses its sign. Figure \ref{fig:octtrunc} shows this in
numerical integrations of the DA equations, truncated at the octupole and
through the hexadecapole, for two example values of $C_K$ on either side
of $C_K^{*}$. Below $C_K^{*}$ (top panel) the hexadecapole makes the slow
oscillation of $\Jh$ about four times smaller and its period about five
times longer; above $C_K^{*}$ (bottom panel) $\Jh$ no longer crosses
zero.
The near cancellation is a property of $\Lambda$ as a whole.
$\Lambda=\chi_\text{oct}\left(\chi_\text{oct}F_\text{lib}/2+c_\delta\right)
+4\chi_2$ (Equation 19 of \citepaperIV, corrected) has two parts: the
correlation of each kick with the phase slip the previous kick
produced, $c_\delta$ being the first-order octupole phase slip, and the
second-order kick harmonic $\chi_2$ (both in Equation 15 of
\citepaperIV). Across the librating window each part changes by several
times $-4\chiH^\text{lib}$, in opposite directions, while their sum
stays close to it. There is no exact relation between $\Lambda$ and
$\chiH^\text{lib}$: their ratio is exactly $16/17$ at the deep edge but
diverges at the separatrix. Why it stays within about 2 per cent of $16/17$
from the deep edge to $C_K=-0.5$ is an open question.

\begin{figure}
 \begin{centering}
 \includegraphics[width=0.92\columnwidth]{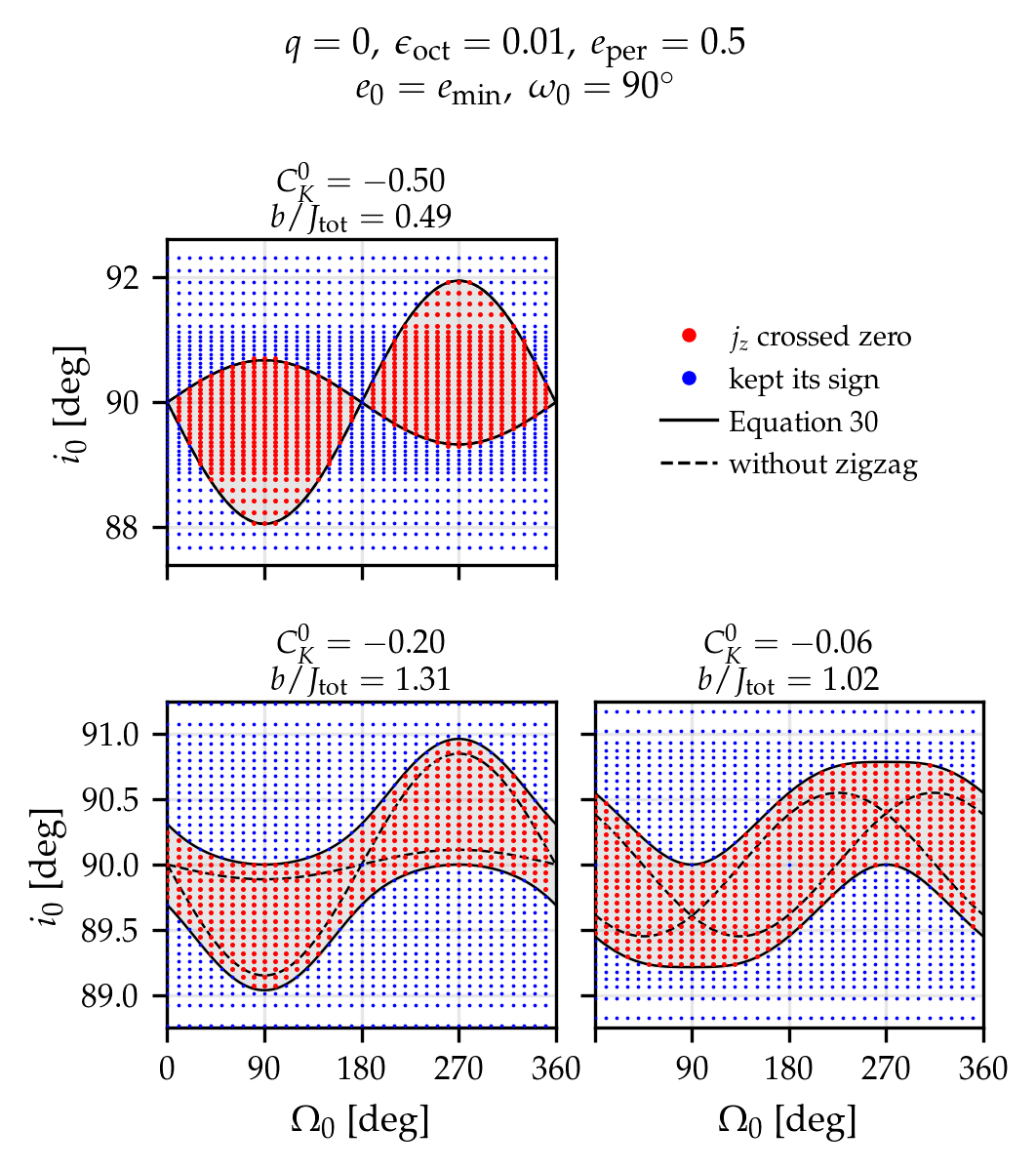}
 \par\end{centering}
 \caption{The flip window in the observable
 $\left(\Omega_0,i_0\right)$ plane, in numerical integrations of the DA
 equations through the hexadecapole, one value of $C_K^0$ per panel,
 with the shared parameters and initial conditions above the panels.
 Red points
 mark integrations in which $j_z$ crossed zero and blue points those
 in which it kept its sign. The gray lens bounded by solid curves is
 the criterion of Equation \ref{eq:flipq}. The dashed curves are the
 same criterion without the zigzag terms, the flips of the slow
 momentum alone.
 \label{fig:flipq0}}
\end{figure}

\begin{figure}
 \begin{centering}
 \includegraphics[width=0.92\columnwidth]{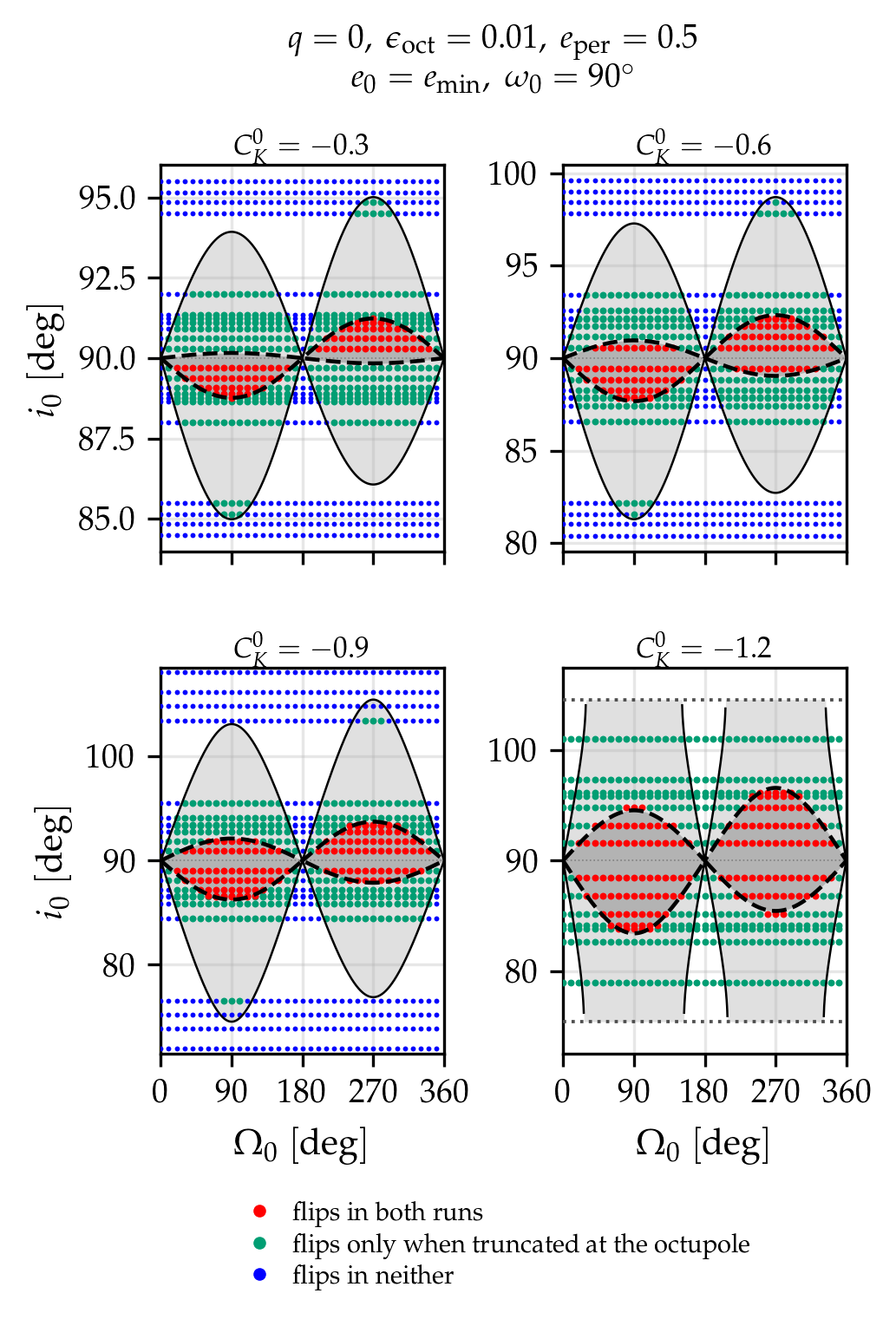}
 \par\end{centering}
 \caption{The flip window of Figure 5 of \citepaperIV\ in the observable
 $\left(\Omega_0,i_0\right)$ plane, in numerical integrations of the DA
 equations truncated at the octupole and through the hexadecapole, one
 value of $C_K^0$ per panel, with the shared parameters and initial
 conditions above the panels. Red points mark initial conditions that
 flip in both integrations, green points those that flip only in the
 octupole-truncated one, and blue points those that flip in neither.
 The light gray lens bounded by solid curves is the flip criterion of
 \citepaperIV, and the darker lens bounded by dashed curves is
 Equation \ref{eq:flipq}. The rows are those of \citepaperIV,
 $j_{z,0}=\pm\left\{0.25,0.4,0.9,0.97,1.03,1.1\right\}$ times its
 threshold, and rows at $\pm\left\{0.25,0.5,0.75,0.9,1.1\right\}$
 times the threshold of Equation \ref{eq:flipq}. At $C_K^0=-1.2$ the
 dotted lines bound the band in which librating cycles exist, which
 excludes the outer rows of \citepaperIV.
 \label{fig:flipplaneIV}}
\end{figure}

At $q=1$ an orbit flips if and only if the slow momentum crosses zero
(Section \ref{sec:flips}). For $q<1$ the zigzag lets $j_z$ cross zero
while $\Jh$ keeps its sign (bottom panel of Figure \ref{fig:octtrunc}). In
\citepaperIV this cannot happen, since there the zigzag is narrower
than the separatrix across the librating window, $b<J_\text{tot}$ with
$J_\text{tot}\equiv\sqrt{2\left|C_\text{tot}/F_\text{lib}\right|}$, which reduces
to $J_\text{env}$ at $q=1$. The near cancellation reverses this
inequality near $C_K^{*}$ and $q^{*}$. Conservation of
$\frac{F_\text{lib}}{2}\Jh^2-\frac{C_\text{tot}}{2}\cos2\psi$ bounds
$\left|\Jh\right|$ along each slow trajectory, and comparing this
bound with the zigzag gives the flip criterion
\begin{equation}
\Jh_0^2<\left\{\begin{array}{@{}l@{\ \ }l@{}}
J_\text{tot}^2\sin^2\Omega_0+\max(0,b^2-J_\text{tot}^2)
& C_\text{tot}>0,\\
J_\text{tot}^2\cos^2\Omega_0+b^2 & C_\text{tot}<0,
\end{array}\right.
\label{eq:flipq}
\end{equation}
with $\Jh_0=j_{z,0}-b\sin\Omega_0$ as above. For $C_\text{tot}>0$
and $b<J_\text{tot}$ this is the criterion of \citepaperIV with
$C_\text{tot}$ in place of the octupole torque. For $C_\text{tot}<0$
the zigzag always widens the flip window, because circulating
trajectories pass closest to $\Jh=0$ at $\psi=\pm\pi/2$, where the
zigzag is largest. Figure \ref{fig:flipq0} shows the flips in numerical integrations at
$q=0$ for three values of $C_K^0$, one per branch of the criterion,
with Equation \ref{eq:flipq} (solid) and the same criterion without the
zigzag terms (dashed). At $C_K^0=-0.5$, where $C_\text{tot}>0$ and
$b<J_\text{tot}$, the flips fill a lens pinched at $\Omega_0=0^\circ$
and $180^\circ$, as in \citepaperIV, and the zigzag terms leave it
unchanged. At $C_K^0=-0.2$, where the near cancellation makes
$b>J_\text{tot}$, and at $C_K^0=-0.06$, above $C_K^{*}$, the flips
instead fill a band about $i_0=90^\circ$ without pinches, wider than
the window of the slow momentum alone. In each panel the flips fill
the window of Equation \ref{eq:flipq}.

Applied to the flip windows of \citepaperIV, Equation \ref{eq:flipq}
narrows the lens by about a factor of four in $j_{z,0}$ at $q=0$.
Figure \ref{fig:flipplaneIV} repeats the integrations of Figure 5 of
\citepaperIV\ ($\epsilon_\text{oct}=10^{-2}$), from the same initial
conditions, through the hexadecapole, with $\eper=0.5$, a choice that
leaves Equation \ref{eq:lameff} unchanged, since
$\epstwo=\epsilon^2_\text{oct}$ at $q=0$
(Equation \ref{eq:bookkeeping}). Through the hexadecapole most of the
initial conditions that flip in the octupole-truncated integrations no
longer flip, and none flips only through the hexadecapole; the flips
that remain fill the lens of Equation \ref{eq:flipq}. At
$C_K^0=-1.2$ the octupole-truncated lens dissolves, as in
\citepaperIV, while the lens through the hexadecapole does not,
consistent with its tips, at $j_{z,0}\simeq0.049$, lying below the
fixed-point ceiling $J_\text{ceil}\simeq0.08$
(Section \ref{sec:landscape}), whereas those of \citepaperIV, at
$0.18$, lie above it.

\needspace{4\baselineskip}
\section{Discussion}
\label{sec:discussion}

In this Letter we extended the analytic description of the
high-eccentricity secular dynamics in Papers I--IV to the hexadecapole.
The leading slow equations reduce to a simple pendulum in
$2\Omega_e$ on both KLC classes, with closed-form hexadecapole
coefficients. For equal inner masses this is the leading dynamics
that changes $j_z$. For unequal masses the octupole generally
controls the rotating class, while on the librating class its
second-order torque must be combined with the hexadecapole, since the
two are of the same hierarchy order, $O\left(\alpha^2\right)$, and
depend on the same angle, $2\Omega_e$. In the
limit $q\to0$ the two nearly cancel, and their sum changes sign at
$C_K^{*}\simeq-0.146$ (Equation \ref{eq:ckstar}). This is contrary to
the expectation that for $q\to0$, where the octupole is strongest,
truncating at the octupole suffices.

The hexadecapole also sets how large the librating excursions are.
Truncated at the octupole, they are comparable to those on the rotating
class at $\epsilon_\text{oct}\gtrsim3\times10^{-2}$, but are reached on a
longer timescale (\citepaperIV). Through the hexadecapole, the near
cancellation removes most of the slow torque that drives them as
$q\to0$. At $q=0$ in Figure \ref{fig:qsweep}
($\epsilon_\text{oct}=0.017$), the largest librating excursion is
$0.05$, against $0.17$ in integrations from the same initial conditions
truncated at the octupole, and $0.29$ on the rotating class. For small
$q$, librating excursions are thus several times smaller than rotating
ones.

Whether an initial condition starts on the librating class depends
only on its inclination and argument of pericenter: by
Equation \ref{eq:CK}, $C_K<0$ when $\sin^2i_0\sin^2\omega_0>2/5$.
For isotropic orientations and
uniform $\omega_0$, $\left|\sin i_0\sin\omega_0\right|$ is uniform on
$[0,1]$, giving a fraction $1-\sqrt{2/5}\simeq36.8$ per cent. The
octupole raises $C_K$ by up to about $2\epsilon_\text{oct}$, so a
system remains on this class only if $C_K^0\lesssim-2\epsilon_\text{oct}$.
Since
$\left|C_K^0\right|\leq\frac{3}{2}e_0^2$, near-circular initial
conditions cross into the rotating class within the first cycles,
where the octupole dominates. Planetary ensembles, started near-circular
($e_0=0.01$ in most runs of \citealt{naoz2012}) or with Rayleigh
distributions whose mean must be about 0.1--0.2 to match the observed
cold Jupiters \citep{weldon2025}, lie mostly in this regime. Stellar
inner orbits are more eccentric: \citet{shariat2026frb} draw the inner
eccentricities from the distributions of \citet{moe2017}, and
\citet{shariat2023wd,shariat2025cv,shariat2025lmxb} from a uniform one,
both with isotropic inclinations. For eccentricity distributions
$p\left(e_0\right)\propto e_0^{\eta}$ with $0\leq\eta\leq1$, which
bracket these, a quarter to a third of
all initial conditions remain librating for
$\epsilon_\text{oct}=0.01$--$0.02$. On these, while the outer orbit
precesses slowly (Section \ref{sec:setup}), the combined slow torque has
the hexadecapole sign, which an integration truncated at the octupole
misses: at every $C_K<0$ for $q>0.06$, since $q^*<0.06$
(Section \ref{sec:liboct}), and above $C_K^{*}$ for smaller $q$.
Most of the stellar triples above do not meet that condition: their
tertiaries are mostly lighter than the inner binary, which then
precesses the outer pericenter faster than the slow evolution proceeds.

\begin{figure}[t]
 \begin{centering}
 \includegraphics[width=0.92\columnwidth]{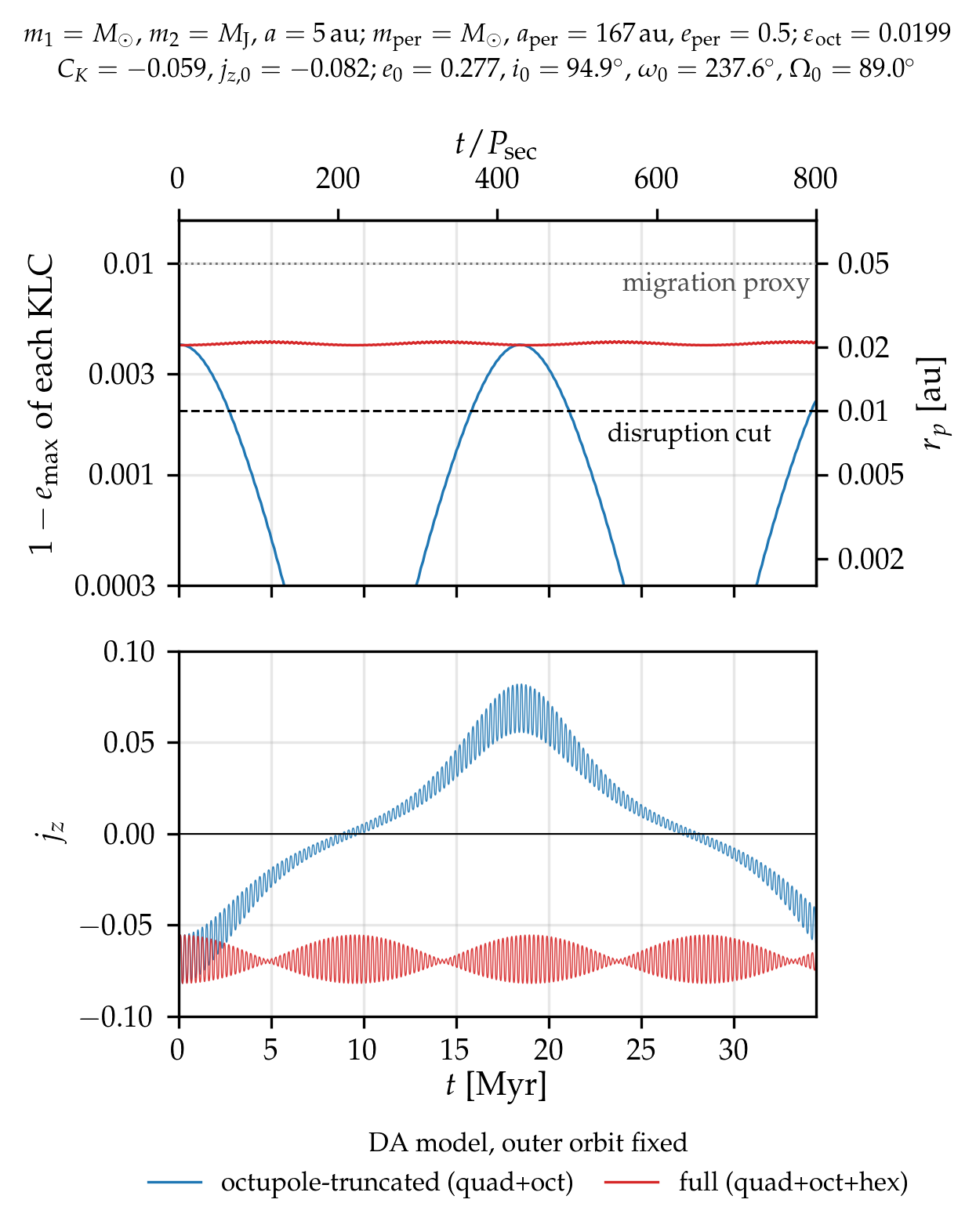}
 \par\end{centering}
 \caption{The planetary example: a Jupiter around a solar-mass star
 with a solar-mass companion, with the parameters and initial
 conditions above the plot. The DA equations are integrated with the
 outer orbit fixed, truncated at the octupole (blue) and through the
 hexadecapole (red). Top panel: $1-e_{\max}$ of every KLC, equal
 to $r_p/a$, with the disruption cut ($r_p=0.01\,$au) and the
 migration proxy ($r_p<0.05\,$au) as reference lines. Bottom panel:
 $j_z$.
 \label{fig:appetizer}}
\end{figure}

Population calculations often integrate the DA equations truncated at
the octupole order (Section \ref{sec:intro}), treating the hexadecapole
as subdominant for wide outer orbits and unequal inner masses
\citep[e.g.,][]{shariat2026frb}. Within the assumptions of this Letter,
these do not alone make it negligible: the two librating
torques have the same hierarchy scaling. Both must be retained to obtain the
leading slow evolution.

The analysis in this Letter assumes $\left|j_z\right|\ll1$, a fixed
outer orbit, and slow evolution relative to the KLC. The coefficients
of the slow equations are computed on the quadrupole cycles at $j_z=0$
(Section \ref{sec:cycles}) and at leading order in $\alpha$; near a zero
of a coefficient the terms of higher order in $\alpha$ become
important, and near the fixed-point ceiling
(Section \ref{sec:landscape}) the dependence on $j_z$.
General relativity (GR), tides and rotation can also change the cycles
and their slow evolution. Two further effects act on the slow pendulum
directly: the precession of the outer pericenter
(Section \ref{sec:setup}), and the precession of $\Omega_e$ added by
Brown's Hamiltonian, of relative order $P_\text{out}/P_\text{sec}$, where
$P_\text{out}$ is the outer orbital period. Each shifts the center of
the pendulum away from $j_z=0$, as Brown's Hamiltonian does on rotating
cycles (\citepaperII), and the results here require that shift to be
small compared with the width of the pendulum. For equal inner masses on
librating cycles, a heavier perturber slows the first precession but
speeds up the second, so at $\eper=0.5$ no perturber mass keeps both
shifts small, and the librating results of Section \ref{sec:equalmass}
describe the fixed-orbit DA model rather than a realizable triple. The fraction of
librating initial conditions therefore does not by itself give
an error in a population's merger or disruption rate.

An example is shown in Figure \ref{fig:appetizer}: a Jupiter at
$5\,$au around a solar-mass star, with a solar-mass companion at
$167\,$au on an $\eper=0.5$ orbit. The octupole-truncated DA
integration reaches the adopted disruption cut, $r_p=0.01\,$au, at
$2.8\,$Myr and, if continued, flips within $10\,$Myr. Including the
hexadecapole, the minimum pericenter remains above $0.020\,$au and the
orbit does not flip during the $1\,$Gyr integration. The example lies
in the librating regime above $C_K^{*}$ ($C_K=-0.06$), where the
combined slow torque has the opposite sign to the octupole contribution
(Section \ref{sec:liboct}).

Adding short-range forces (SRFs) such as GR and the planet's
conservative tidal bulge, separately or together, leaves the
full-model minimum pericenter above $0.020\,$au. In the truncated model, GR and the tidal bulge together
limit the minimum to $0.0098\,$au, half the full-model value. The inner
semimajor axis is held fixed, with $r_p<0.05\,$au as a proxy for tidal
migration, which is not modeled; both models enter that regime, at
$r_p\simeq0.02\,$au, from the first cycle. The example therefore
compares the eccentricities the two models drive, not the fate of the
planet.

Where the DA approximation breaks down, the hexadecapole can become
necessary for a further reason. There, SRFs drive
discrete, nonadiabatic jumps of the quadrupole invariants, $C_K$ among
them, during high-eccentricity episodes
\citep{hamilton2024,kleinhamilton2026a}. With the octupole, an orbit
that starts on the rotating class ($C_K>0$) can be driven to
$j_z\to0$, where such jumps are triggered
\citep{klein_hamilton_in_prep}, and a jump can carry $C_K$ to negative
values, onto the librating class. This holds even for near-circular
initial conditions, with $\left|C_K^0\right|\ll1$: in the example of
\citet{kleinhamilton2026a}, started at $e_0=0$ ($C_K=10^{-6}$), the
jumps carry $C_K$ across almost the entire interval
$\left[-3/2,1\right]$. On the librating class the octupole slow
torque is of second order in $\epsilon_\text{oct}$ (\citepaperIV), the
hexadecapole torque, of first order in $\epstwo$, competes with it, and
for $q\to0$ the two nearly cancel (Section \ref{sec:liboct}). Wherever
these jumps matter, so that the single-averaged equations are needed,
the hexadecapole is therefore needed as well.

\begin{acknowledgments}
We thank Pavel Ivanov for the stimulating question.
\end{acknowledgments}

\bibliographystyle{aasjournal}
\bibliography{hexadecapole_nearly_cancels_the_octupole}

\restartappendixnumbering
\appendix
\renewcommand{\theHfigure}{\thesection\arabic{figure}}
\renewcommand{\theHtable}{\thesection\arabic{table}}

\section{Circular perturber}
\label{sec:circular}

For a circular perturber, $\eper=0$, the octupole and the
nonaxisymmetric hexadecapole vanish. The surviving DA interaction
is axisymmetric at every multipole order, so $j_z$ is conserved and
orbital flips are absent. The axisymmetric hexadecapole changes the
KLCs through $f_0$ (Section \ref{sec:setup}), and the modified cycles
still reduce to a single integral over $e_z$, as shown below. $C_K$ is no longer
conserved. For the quadrupole, $C_K$ is $\frac{4}{3}\phi_\text{quad}$
without its constant and its term in $j_z$ alone
(Equation \ref{eq:phi_quad}); the same operation on $\phi$ gives a
conserved $\tilde{C}_K$ (Equation \ref{eq:CKtilde}), which reduces to
$C_K$ at $\epsh=0$. For $\epsh<64/165$ its sign separates the librating
($\tilde{C}_K<0$) from the rotating ($\tilde{C}_K>0$) cycles:
$\tilde{C}_K\ge0$ wherever $e_z=0$ (Equation \ref{eq:CKcross}), so a
cycle with $\tilde{C}_K<0$ never crosses $e_z=0$, and every cycle with
$\tilde{C}_K>0$ does. The bound is sharp: above it, the radial state
$\mathbf{e}=\hat{\mathbf{z}}$, $\mathbf{j}=0$, an equilibrium, has
$\tilde{C}_K=-\frac{3}{2}+\frac{495}{128}\epsh>0$, and the cycles near it
never cross $e_z=0$.
With $j_z$, $\tilde{C}_K$ fixes the cycle up to its azimuth and, on the
librating class, the sign of $e_z$.

We use the vectors $\boldsymbol\alpha=\mathbf{j}+\mathbf{e}$ and
$\boldsymbol\beta=\mathbf{j}-\mathbf{e}$, which satisfy
$\alpha^2=\beta^2=1$.\footnote{They go back to \citet{souriau1969}
and have been used in secular dynamics by \citet{boue2014} and
\citet{tremaine2023book}.} \citet{basha2025} showed with them that the
quadrupole KLC is a simple pendulum at every $j_z$. Since $w\equiv\alpha_z+\beta_z=2j_z$ is constant here, the
orbit, apart from its azimuth, is described by
$v\equiv\alpha_z-\beta_z=2e_z$ and by
\begin{equation}
A\equiv\alpha_x\beta_x+\alpha_y\beta_y,\qquad
B\equiv\left(\boldsymbol\alpha\times\boldsymbol\beta\right)_z=2s,
\label{eq:AB}
\end{equation}
with $s\equiv\left(\mathbf{e}\times\mathbf{j}\right)_z$ as in
Section \ref{sec:kick}. With $j^2+e^2=1$, the eccentricity follows from
$A$ and $v$,
\begin{equation}
e^{2}=\frac{1}{2}\Big[1-A+\frac{1}{4}\left(v^{2}-w^{2}\right)\Big].
\label{eq:e2Av}
\end{equation}
The variables $A$, $B$ and $v$ are related by
\begin{equation}
A^2+B^2=L\left(v\right)\equiv
\Big(1-\frac{\left(v+w\right)^2}{4}\Big)
\Big(1-\frac{\left(v-w\right)^2}{4}\Big).
\label{eq:Lv}
\end{equation}
The orbit therefore has a single degree of freedom. For a circular
perturber this holds at every multipole order, since all are
axisymmetric, and the reduction to a single integral carried out below
for $f_0$ can be repeated for any order of the expansion. Both $\phi_\text{quad}$ and $f_0$ depend on $B$ only
through $B^2$ ($f_0$ contains $s^2$; Table \ref{tab:hex}), so
after $B^2$ is eliminated with Equation \ref{eq:Lv}, $\phi$ is a
function of $v$ and $A$ alone, and
Equations \ref{eq:milankovitch_j}--\ref{eq:milankovitch_e} give
\begin{equation}
\dot{v}=4B\,\frac{\partial\phi}{\partial A},\qquad
\dot{A}=-4B\,\frac{\partial\phi}{\partial v}.
\label{eq:vA}
\end{equation}
The orbit thus follows a curve of constant $\phi$ in the
$\left(v,A\right)$ plane, on which $B$ is given by
Equation \ref{eq:Lv} up to its sign.

The quadrupole is linear in $A$,
$-\phi_\text{quad}=\frac{3}{8}\left(A+v^{2}\right)-\frac{1}{4}$, and
Equations \ref{eq:vA} then describe the simple pendulum of
\citet{basha2025}. The axisymmetric hexadecapole is quadratic in $A$
(through $e^4$ and $s^2$), so that
\begin{gather}
-\phi=-\frac{45}{128}\epsh\,A^{2}+Q\left(v\right)A+P\left(v\right),\nonumber\\
Q=\frac{3}{8}-\frac{45}{1024}\epsh\left(31v^{2}-w^{2}-12\right),\nonumber\\
P=\frac{3}{8}v^{2}-\frac{1}{4}-\frac{9}{8192}\epsh
\left(405v^{4}-130v^{2}w^{2}-1000v^{2}\right.\nonumber\\
\left.\qquad\qquad+\,5w^{4}-40w^{2}+208\right),
\label{eq:phiA}
\end{gather}
a quadratic in $A$ whose coefficients $Q$ and $P$ are polynomials in
$v$. On a curve of constant $-\phi\equiv\mathcal{E}$, Equation
\ref{eq:phiA} is therefore a quadratic equation for $A$. For
$\epsh<8/105$, its two solutions differ at every point of the orbit, and
one of them lies outside the range $A^2\le L$ allowed by
Equation \ref{eq:Lv}; the orbit follows the other,
\begin{equation}
\begin{gathered}
A\left(v\right)=\frac{2\left(\mathcal{E}-P\right)}{Q+\sqrt{D}},\\
D\left(v\right)=Q^{2}-\frac{45}{32}\epsh\left(\mathcal{E}-P\right),
\end{gathered}
\label{eq:Aroot}
\end{equation}
which reduces to
$A=\frac{8}{3}\left(\mathcal{E}+\frac{1}{4}\right)-v^{2}$ at
$\epsh\rightarrow0$. From Equation \ref{eq:phiA},
$\partial\phi/\partial A=\frac{45}{64}\epsh A-Q$, which on the orbit,
where $A$ is given by Equation \ref{eq:Aroot}, equals $-\sqrt{D}$. In
$\partial\phi/\partial A$, $\phi_\text{quad}$ contributes the constant
$-\frac{3}{8}$, and $\epsh f_0$ the terms
$\epsh\big[\frac{45}{64}A+\frac{45}{1024}\left(31v^{2}-w^{2}-12\right)\big]$,
which are largest at $\mathbf{j}=0$, $\mathbf{e}=\pm\hat{\mathbf{z}}$
($v=\pm2$, $w=A=0$), where they equal $\frac{315}{64}\epsh$ and cancel
$-\frac{3}{8}$ at $\epsh=8/105$. Hence, for $\epsh<8/105$, which we
assume from here on, $\partial\phi/\partial A<0$ at every state, so that
every state lies on
the solution of Equation \ref{eq:Aroot} and
$D=\left(\partial\phi/\partial A\right)^{2}>0$. The first of Equations \ref{eq:vA} then gives $\dot{v}=-4\sqrt{D}\,B$, and with
$B^{2}=L-A^{2}$,
\begin{equation}
\dot{v}^{2}=R\left(v\right)
=16\,D\left(v\right)
\left[L\left(v\right)-A^{2}\left(v\right)\right].
\label{eq:quadrature}
\end{equation}
The variable $v$ oscillates between its extremes $v_{-}=2e_{z,\min}$
and $v_{+}=2e_{z,\max}$, the roots of $L=A^{2}$ (where $B=0$), and the
time from $v_{-}$ to $v$ is the single integral
$\int_{v_{-}}^{v}\mathrm{d}v'/\sqrt{R\left(v'\right)}$.
Because $R$ is even in $v$, the librating
oscillation lies on $0<v_{-}<v_{+}$ or on its reflection to negative
$v$, on which $e_z$ keeps its sign, and the rotating oscillation on
$v_{-}=-v_{+}$, on which $e_z$ passes through zero.

The extremes of $e$ follow without integration. At $v_{\mp}$, $B=0$,
so $A^2=L$, and $-\phi=\mathcal{E}$ in
Equation \ref{eq:phiA} gives
$A=\big(\mathcal{E}-P+\frac{45}{128}\epsh L\big)/Q$, so that
\begin{equation}
Q^{2}L=\Big(\mathcal{E}-P+\frac{45}{128}\epsh L\Big)^{2},
\label{eq:turning}
\end{equation}
a quartic in $v^2$, which at $\epsh=0$ reduces to a quadratic, as for the
quadrupole \citep{basha2025}. Its roots in the range
$0\le v^{2}\le\left(2-|w|\right)^{2}$ allowed by
${|\alpha_z|,|\beta_z|\le1}$ are $v_{\mp}^{2}$, and give the extreme
eccentricities through Equation \ref{eq:e2Av}:
$e_{\min}$ and $e_{\max}$ on the librating oscillation, and $e_{\max}$ on
the rotating one, whose $e_{\min}$, at $v=0$, follows from
Equation \ref{eq:CKcross}.

\begin{figure}[t]
 \begin{centering}
 \includegraphics[width=0.92\columnwidth]{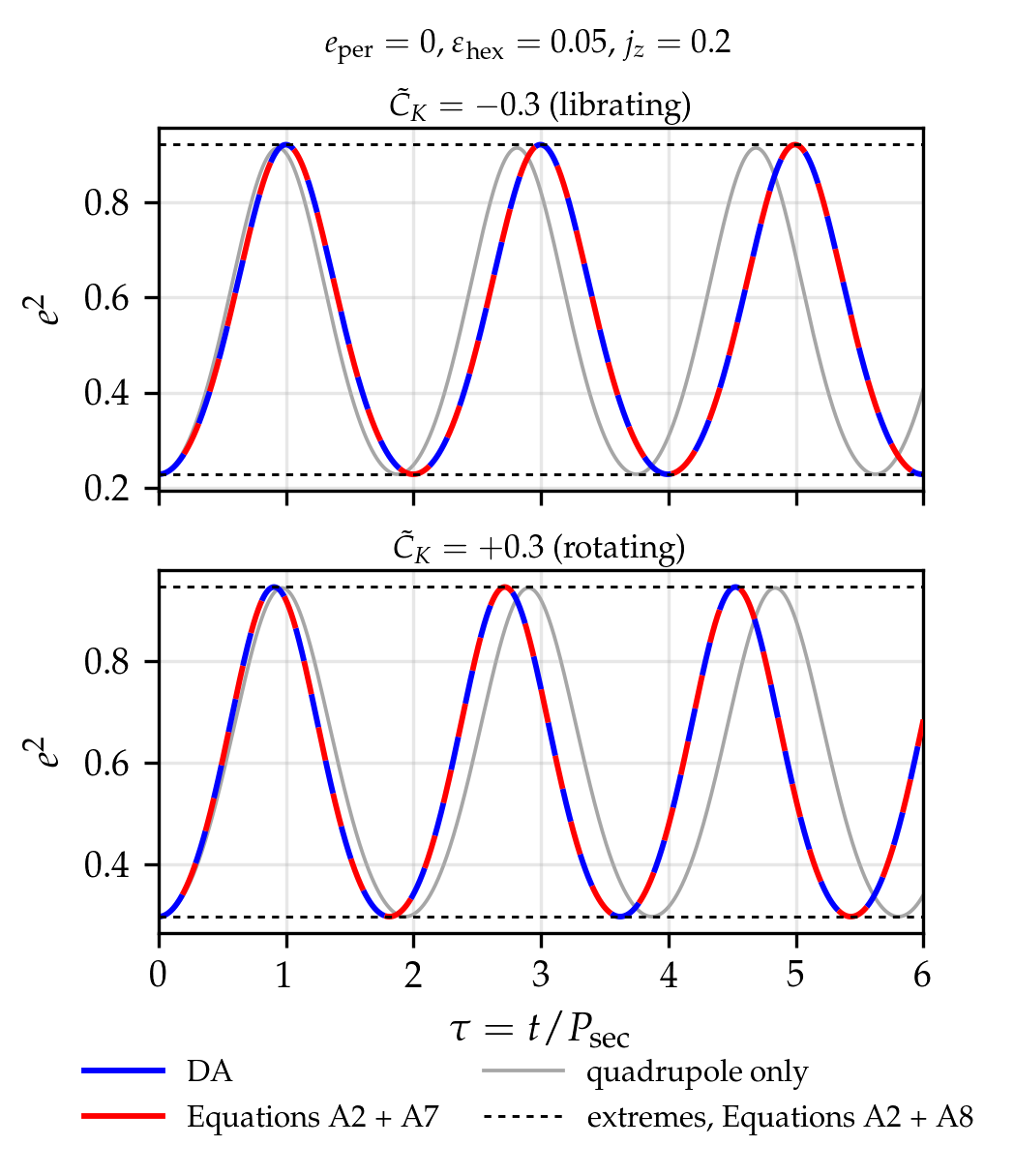}
 \par\end{centering}
 \caption{$e^2$ versus $\tau$ for a librating (top) and a rotating
 (bottom) cycle of a circular perturber, with the parameters above the
 panels: numerical DA integrations (blue) and $e^2$ from the solution
 of Equation \ref{eq:quadrature} through Equation \ref{eq:e2Av} (red), in
 interleaved dashes, the quadrupole-only cycle from the same initial
 conditions (gray), and the extremes of $e^2$ from
 Equations \ref{eq:e2Av} and \ref{eq:turning} (black dashed; on the
 rotating cycle, $e_{\min}$ at $v=0$).
 \label{fig:circular}}
\end{figure}

Equations \ref{eq:Aroot}--\ref{eq:turning} are exact.
Figure \ref{fig:circular} shows, for a librating and a
rotating cycle, $e^2$ (Equation \ref{eq:e2Av}) from the solution of
Equation \ref{eq:quadrature},
over-plotted on numerical DA integrations, with which it coincides.
Relative to the quadrupole-only cycles from the same initial states
(gray), the axisymmetric hexadecapole slightly changes the period and
$e_{\max}$.

\needspace{4\baselineskip}
\section{Hexadecapole coefficients}
\label{sec:coefficients}

Table \ref{tab:hex} lists the polynomials $f_0$ and $f_2$ of
the DA hexadecapole (Equation \ref{eq:phihex}). The coefficients of the
torque in Equation \ref{eq:djzhex} are
\begin{align}
A_2&=\frac{45}{1024}\Big[
28e^2\sin^2 i_e\left(1-4e^2+21e^2\cos^2 i_e-3j_z^2\right)
\nonumber\\
&\qquad+\Big(\frac{j_z^2\cos^2 i_e}{\sin^2 i_e}
-\frac{s^2}{e^2\sin^2 i_e}\Big)G
\nonumber\\
&\qquad+197\,e^2j_z^2\cos^2 i_e+27\,s^2\Big],
\nonumber\\
B_2&=-\frac{45}{1024}\,j_z\,s\cos i_e\left[
\frac{2G}{e\sin^2 i_e}+170\,e\right],
\label{eq:A2B2}
\end{align}
with
\begin{equation}
G\equiv59e^2-111e^2\cos^2 i_e+28j_z^2-4
\label{eq:G}
\end{equation}
and the angles as in Equation \ref{eq:evec}. On the exact cycles
($j_z=0$), the state rotated to $\Omega_e=0$ has
$\mathbf{e}=\left(e_x,0,e_z\right)$ and $\mathbf{j}=\left(0,j_y,0\right)$,
and there
\begin{align}
1024\,f_2\left(\Omega_e{=}0\right)=&-630\,e_x^2-90\,j_y^2+2520\,e_x^2e^2
\nonumber\\
&-13230\,e_x^2e_z^2+\frac{2655}{2}\,j_y^2e^2
\nonumber\\
&-\frac{4995}{2}\,j_y^2e_z^2-\frac{1215}{2}\,e_x^2j_y^2,
\label{eq:f2zero}
\end{align}
with $e^2=e_x^2+e_z^2$, so that $A_2=-2f_2\left(\Omega_e{=}0\right)$ there.

For the circular perturber of Appendix \ref{sec:circular}, removing from
$\phi=\phi_\text{quad}+\epsh f_0$ its constant terms and its terms in
$j_z$ alone, and multiplying by $\frac{4}{3}$, gives
\begin{equation}
\begin{aligned}
\tilde{C}_K={}&e^2-\frac{5}{2}e_z^2+\frac{5}{128}\epsh\big(85e^2-55e_z^2\\
&-49e^4+441e_z^4-323e^2e_z^2\\
&-37e^2j_z^2-294e_z^2j_z^2-97s^2\big).
\end{aligned}
\label{eq:CKtilde}
\end{equation}
Where $e_z=0$, $\mathbf{j}\perp\mathbf{e}$ gives
$s^2=e^2\left(1-e^2-j_z^2\right)$, and Equation \ref{eq:CKtilde}
reduces to
\begin{equation}
\tilde{C}_K=e^2\Big[1+\frac{15}{32}\epsh\left(4e^2+5j_z^2-1\right)\Big],
\label{eq:CKcross}
\end{equation}
which is positive for $e>0$ and $\epsh<32/15$.

\begin{table}
\centering
\small
\caption{The exact DA hexadecapole (Equation \ref{eq:phihex}):
$1024\,f_0$ (axisymmetric) and $1024\,f_2$ (nonaxisymmetric, with the
outer pericenter along $\hat{\mathbf{x}}$). Each pair of columns lists
monomials and their coefficients, with
$s\equiv\left(\mathbf{e}\times\mathbf{j}\right)_z$,
$u_e=e_x^2-e_y^2$, $u_j=j_x^2-j_y^2$,
$w_1=e_zj_z\left(e_xj_x-e_yj_y\right)$ and
$w_3=s\left(e_xj_y+e_yj_x\right)$.}
\label{tab:hex}
\begin{tabular}{lrlr}
\toprule
\multicolumn{2}{c}{$1024\,f_0$} & \multicolumn{2}{c}{$1024\,f_2$}\\
\midrule
$1$        & $54$    & $u_e$        & $-630$\\
$e^2$      & $2550$  & $u_j$        & $90$\\
$e_z^2$    & $-1650$ & $u_e e^2$    & $2520$\\
$j_z^2$    & $-540$  & $u_e e_z^2$  & $-13230$\\
$e^4$      & $-1470$ & $u_e j_z^2$  & $1890$\\
$e_z^4$    & $13230$ & $u_j e^2$    & $-2655/2$\\
$j_z^4$    & $630$   & $u_j e_z^2$  & $4995/2$\\
$e^2e_z^2$ & $-9690$ & $u_j j_z^2$  & $-630$\\
$e^2j_z^2$ & $-1110$ & $w_1$        & $8865/2$\\
$e_z^2j_z^2$ & $-8820$ & $w_3$      & $-1215/2$\\
$s^2$      & $-2910$ & & \\
\bottomrule
\end{tabular}
\end{table}

\needspace{4\baselineskip}
\section{Orbital angles}
\label{sec:angles}

\begin{figure}
 \begin{centering}
 \includegraphics[width=0.85\columnwidth]{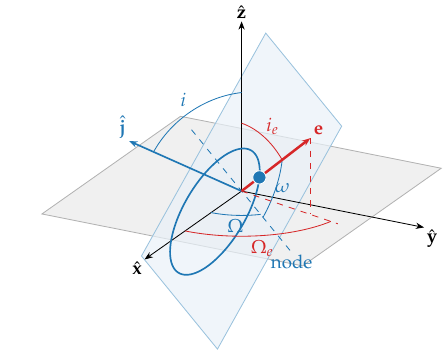}
 \par\end{centering}
 \caption{The frame and the angles. $\hat{\mathbf{z}}$ lies along the
 outer angular momentum and $\hat{\mathbf{x}}$ points toward the outer
 pericenter; the gray parallelogram is the outer orbital plane. The
 inner orbit (blue ellipse) lies in its own plane (light blue), which
 crosses the outer one along the line of nodes; its orbital elements $i$, $\Omega$ and
 $\omega$, and the unit vector $\hat{\mathbf{j}}$ along its angular
 momentum, are drawn in blue. The eccentricity vector $\mathbf{e}$
 points toward the pericenter of the inner orbit (dot); its polar angle
 $i_e$ and its
 azimuth $\Omega_e$, measured from $\hat{\mathbf{x}}$ to the projection
 of $\mathbf{e}$ on the outer plane (dashed), are drawn in red. The angles
 shown are $i=60^\circ$, $\Omega=45^\circ$ and $\omega=50^\circ$, giving
 $i_e=48^\circ$ and $\Omega_e=76^\circ$.
 \label{fig:angles}}
\end{figure}

Figure \ref{fig:angles} shows the frame of Section \ref{sec:setup} and
the two sets of angles used in this Letter: the inclination $i$, the
longitude of the ascending node $\Omega$ and the argument of pericenter
$\omega$ of the inner orbit, and the polar angle
$i_e$ and azimuth $\Omega_e$ of its eccentricity vector
(Equation \ref{eq:evec}). They are related by
\begin{equation}
\begin{gathered}
\cos i_e=\sin i\sin\omega,\\
\tan\Omega_e=\frac{\sin\Omega\cos\omega+\cos\Omega\cos i\sin\omega}
{\cos\Omega\cos\omega-\sin\Omega\cos i\sin\omega},
\end{gathered}
\label{eq:angles}
\end{equation}
so that $\Omega_e=\Omega$ for $\omega=0$, and $\Omega_e=\Omega$ modulo
$\pi$ for $j_z=0$ ($i=90^\circ$), where $\mathbf{e}$ moves in the plane
spanned by $\hat{\mathbf{z}}$ and the line of nodes
(Section \ref{sec:flips}).

\section{Excursion profiles and scalings}
\label{sec:profiles}

\begin{figure}
 \begin{centering}
 \includegraphics[width=0.92\columnwidth]{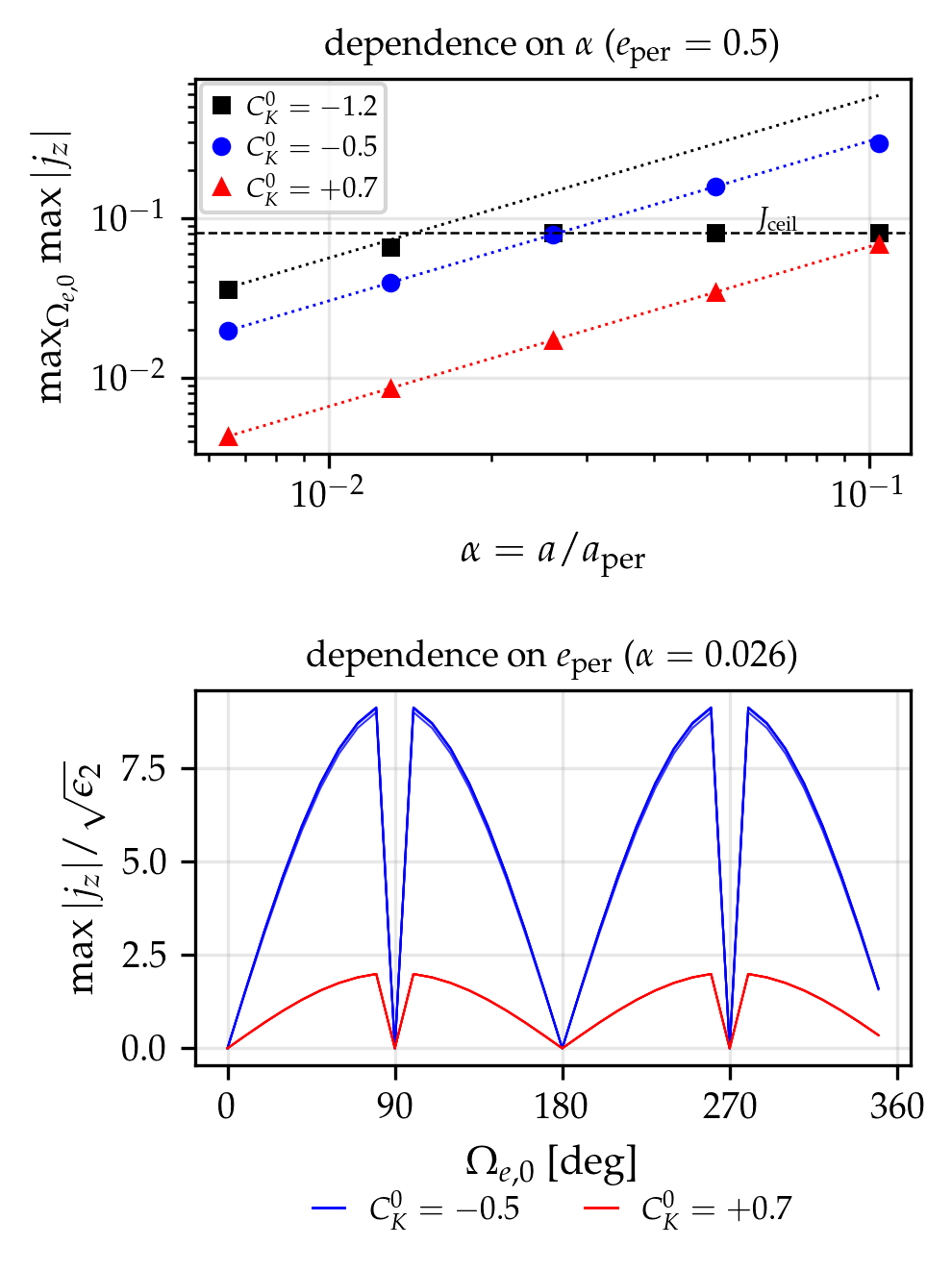}
 \par\end{centering}
 \caption{Dependence of the angular-momentum excursions on the
 hierarchy and on the outer eccentricity ($q=1$). Top panel: the
 maximum over initial azimuths of $\max\left|j_z\right|$ versus $\alpha$
 at $\eper=0.5$, for three values of $C_K^0$, with the frozen
 prediction, which follows $\sqrt{\epstwo}\propto\alpha$ (dotted); at
 $C_K^0=-1.2$ the excursions reach the ceiling $J_\text{ceil}$
 (dashed) for $\alpha\geq0.026$, independent of $\alpha$. Bottom
 panel: the azimuth profiles scaled by $\sqrt{\epstwo}$ at fixed
 $\alpha=0.026$, for $\eper$ from $0.1$ to $0.7$ at $C_K^0=-0.5$ and
 $+0.7$; they collapse onto one curve per value of $C_K^0$.
 \label{fig:scalings}}
\end{figure}

\begin{figure*}
 \begin{centering}
 \includegraphics[width=\textwidth]{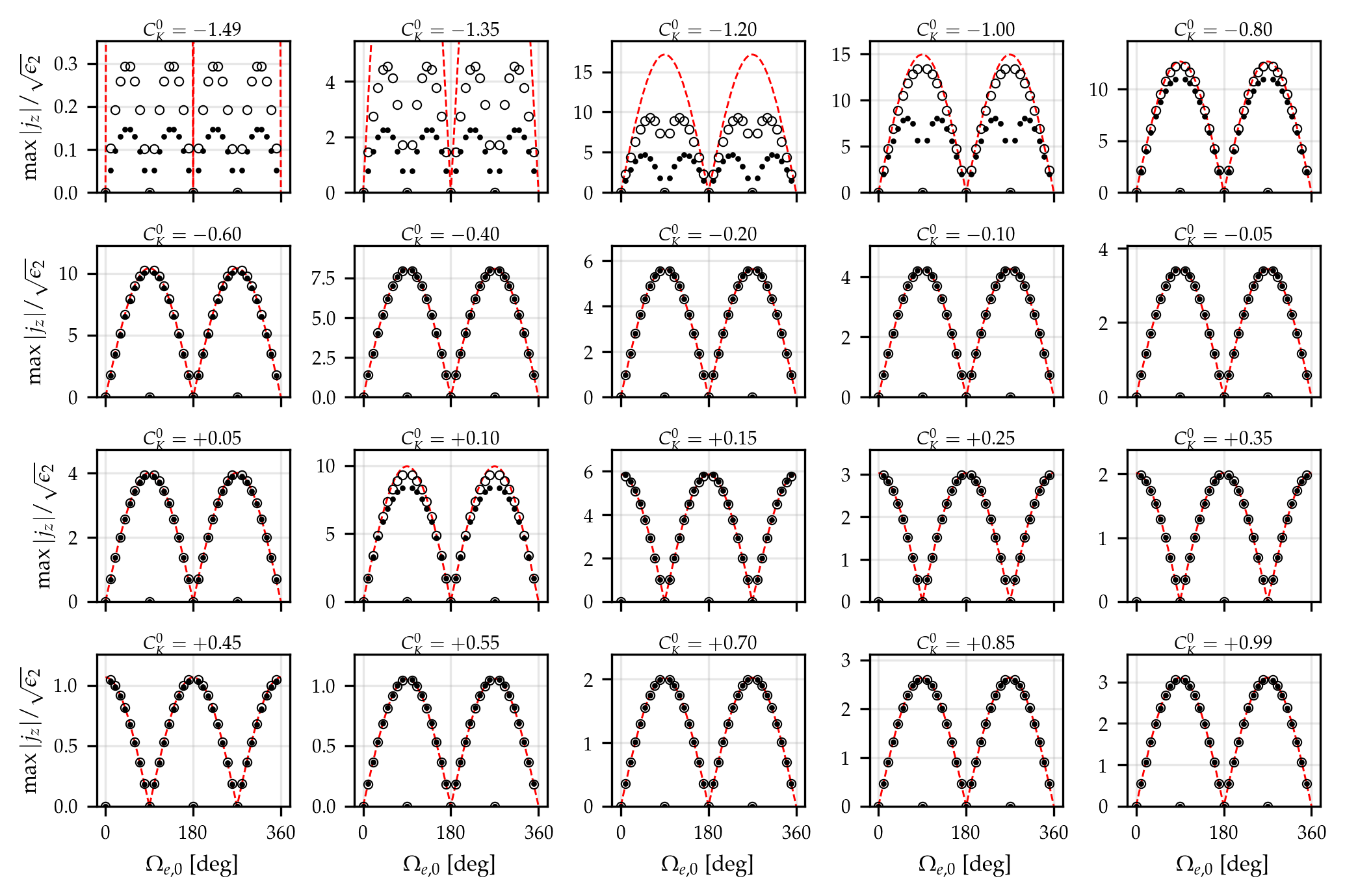}
 \par\end{centering}
 \caption{The scaled angular-momentum excursions ($q=1$, $\eper=0.5$): $\max\left|j_z\right|/\sqrt{\epstwo}$
 from $j_{z,0}=0$ versus the initial azimuth $\Omega_{e,0}$ at twenty values
 of $C_K^0$ spanning both KLC classes, for $\alpha=0.052$
 (black dots) and $\alpha=0.026$ (open circles), with the leading
 pendulum (red dashed; Equation \ref{eq:jmax}). In the deepest panels the frozen
 prediction extends above the plotted range, while the numerical
 excursions approach the ceiling $J_\text{ceil}$ of Figure
 \ref{fig:jzminmax}, which does not scale with $\alpha$.
 \label{fig:landscape}}
\end{figure*}

Figure \ref{fig:scalings} shows how the excursions of the numerical
integrations of Section \ref{sec:landscape} scale with $\alpha$ and
$\eper$, and Figure \ref{fig:landscape} shows them as a function of
the initial azimuth, at each of the twenty values of $C_K^0$.

\end{document}